\documentclass[  
 aps,
]{revtex4}
\usepackage{graphicx}
\usepackage[colorlinks=true, pdfstartview=FitV, linkcolor=blue, citecolor=red, urlcolor=magenta]{hyperref}
\usepackage{graphicx}
\usepackage{latexsym}
\usepackage{amsmath}
\usepackage{amsfonts}
\usepackage{amssymb}
\usepackage{verbatim}
\usepackage{braket}
\usepackage{extarrows}
\usepackage{hyperref}
\usepackage{csquotes}

\begin{document}

\title{Vacuum energy density for interacting real and complex scalar fields in a
Lorentz symmetry violation scenario}
\author{$^{1}$A. J. D. Farias Junior}
\email{antonio.farias@ifal.edu.br}
\author{$^{2}$A. Smirnov}
\email{smirnov@ufs.br}
\author{$^{3}$Herondy F. Santana Mota}
\email{hmota@fisica.ufpb.br}
\author{$^{3}$E. R. Bezerra de Mello}
\email{emello@fisica.ufpb.br}
\affiliation{$^{1}$Instituto Federal de Alagoas,\\
CEP: 57460-000, Piranhas, Alagoas, Brazil.} 
\affiliation{$^{2}$Departamento de F\'{\i}sica, Universidade Federal de Sergipe,\\
CEP 49107-230-, S\~{a}o Cr\'{\i}stov\~{a}o, Sergipe, Brazil.} 
\affiliation{$^{3}$Departamento de F\'{\i}sica, Universidade Federal da Para\'{\i}ba,\\
Caixa Postal 5008, Jo\~{a}o Pessoa, Para\'{\i}ba, Brazil.}

\begin{abstract}
In this paper the vacuum energy density and generation of topological mass
are investigated for a system of a real and complex scalar fields interacting with each other. In addition to that, it is
also included the quartic self-interaction for each one of the fields. The
condition imposed on the real field is the periodic condition, while the
complex field obey a quasi-periodic condition. The system is placed in a
scenario where the CPT-even aether-type Lorentz symmetry violation takes
place. We allow that the Lorentz violation affects the fields with different
intensities. The vacuum energy density, its loop correction, and the topological mass
are evaluated analytically. It is also discussed the possibility of
different vacuum states and their corresponding stability requirements, which
depends on the conditions imposed on the fields, the interaction coupling constants and also the
Lorentz violation parameters. The formalism used here to perform this
investigation is the effective potential one, which is written as a loop
expansion via path integral in quantum field theory.
\end{abstract}

\maketitle

\section{Introduction}

The Casimir effect is one of the most interesting physical phenomenon which
is of pure quantum nature. This effect was predicted by H. Casimir \cite%
{casimir1948attraction} in 1948. The Casimir effect consists in a force of
attraction that arises between two neutral parallel plates, placed in vacuum
near to each other. The theoretical description of such a force lies in the
framework of the electromagnetic quantum field theory. The force of
attraction is due to modifications in the vacuum fluctuations of the
associated field, as a consequence of the imposed boundary conditions on the
plates. The first experimental attempt to detect the Casimir's prediction
dates back to Sparnaay in 1958 \cite{Sparnaay:1958wg}, and ended up not
having the required precision to confirm the phenomenon without doubt. It
was only several decades later that the Casimir effect was confirmed by
several high accuracy experiments \cite{bressi2002measurement,
kim2008anomalies,
lamoreaux1997demonstration,lamoreaux1998erratum,mohideen1998precision,mostepanenko2000new,wei2010results}%
. Currently, it is known that not only the quantum vacuum modifications of
the electromagnetic field can give rise to a force, that is, the phenomenon
is also manifested for instance in the context of the scalar and fermionic
quantum field theory, at least theoretically. Besides, the nontrivial
topology of a given spacetime can also influence the quantum vacuum
fluctuations in such a way that a vacuum force is induced \cite%
{bordag2009advances, milton2001casimir,mostepanenko1997casimir}. In this sense, the basic concept 
of the Casimir effect has been expanded into a broader set of possibilities 
for the quantum vacuum fluctuations of the underlining field to be altered. As a consequence, it is common
in these other scenarios to say that a Casimir-like effect occurs.

For example, the vacuum energy for a massive fermionic field confined
between two points in one spatial dimension, with the MIT bag model boundary
condition imposed, is considered in Ref.~\cite{Saghian:2012zy}, due to the
influence of the nontrivial topology of the global monopole spacetime in
Ref.~\cite{Saharian:2003sp}, and due to a fermionic chain in Ref.~\cite%
{flachi2017sign}. Considering scalar fields, a Casimir-like effect with the
field subjected to a helix-like boundary condition with temperature
corrections is considered in Ref.~\cite{aleixo2021thermal} and subjected to
Robin boundary conditions in Ref.~\cite{romeo2002casimir}. In Ref.~\cite%
{maluf2020casimir} the vacuum energy for a real scalar field and the Elko
neutral spinor field in a field theory at a Lifshitz fixed point is obtained
and in Ref.~\cite{Escobar:2023hzz} it is investigated a Casimir-like effect in the
classical geometry of two parallel conducting plates, in a noncommutative
spacetime within a coherent state approach. A review on the Casimir effect and
its generalizations
can be found in Ref.~\cite{bordag2009advances,
milton2001casimir,mostepanenko1997casimir}.

In its most standard approach, the investigation of Casimir-like effects is
performed assuming that the Lorentz symmetry is preserved. However, attempts
to build a high energy scale theory fails to preserve the Lorentz symmetry,
even locally when we consider gravity. In this sense, there are two
proposals for the Lorentz symmetry violation. One is described in the
context of string theory in Ref.~\cite{Kos}, where non-vanishing vacuum
expectation values of some vector gives rise to preferential directions in
the spacetime. The other proposal is considered in the context of quantum
gravity in Ref.~\cite{hovrava2009quantum}, where different properties of
scales in the space and time are set, which yields an anisotropy between
space and time. In the scenario which the Lorentz symmetry violation is
allowed, the spacetime becomes anisotropic in some direction which includes
the time one, with the anisotropy being determined mathematically by a
constant unit four-vector. The modifications in the spacetime introduced by
the Lorentz symmetry violation, combined with appropriate conditions,
affect the energy modes of the quantum fields and, as a consequence, a
nonzero vacuum force is induced. The Lorentz violation is a topic in which a
great deal of attention has been given in the past years, mainly because it
is an alternative to obtain new physics beyond the standard model. In Ref.~%
\cite{Aj} a Casimir-like effect for a scalar field in a scenario with Lorentz
symmetry violation as consequence of the presence of a constant vector is
investigated together with a helix-like boundary condition. The influence of
a Lorentz violation to the vacuum energy of a massive scalar field at
finite temperature effects and in the presence of an external magnetic field
is investigated in Ref.~\cite{Santos:2022tbq}. In \cite{cruz2020casimir}, an
aether-type Lorentz symmetry violation theoretical model is considered to
investigate a Casimir-like effect and the generation of topological mass
associated with self-interacting massive scalar fields. A Casimir-like effect
in a scenario with Lorentz violation and considering higher order
derivatives was considered in Ref.~\cite{dantas2023bosonic}. Considering the
context of string theory, the Lorentz symmetry violation is studied in the
case of low-energy scale in Refs.~\cite%
{PhysRevD.55.6760,PhysRevD.58.116002,Ani,Carl,Hew,Ber,Kost,Anc,Ber2,Alf,Alf2}%
. Also in the context of string theory and low-energy scale scenarios with
Lorentz violation, the vacuum energy is investigated in Refs. \cite{Obo,
Obo2, Mar,Mar2,Esc}.

Quantum fields found in nature are always interacting with each other. For
that reason the study of the Casimir effect demands models with interactions
so that the description of the physical phenomenon becomes more realistic.
In this context, the first step is to consider a quartic self-interaction in
building the model, which in fact has been done in a variate of works \cite%
{cruz2020casimir, Aj, toms1980symmetry, porfirio2021ground}. On the other
hand, interacting real scalar fields are considered in \cite%
{toms1980interacting}, being one field twisted and the other untwisted, for
the investigation of symmetry breaking and mass generation. Using
non-equilibrium quantum field theory, the quartic interaction between two
scalar real quantum fields is considered in \cite{wang2022particle}, in
order to investigate the phenomenon of particle production from oscillating
scalar backgrounds in a Friedmann-Lema\^{\i}tre-Robertson-Walker universe.
In Ref.~\cite{PhysRevB.77.115409} a thermodynamics Casimir-like effect is
studied considering interacting field theories with zero modes and in Ref.~%
\cite{article} the thermal corrections to the vacuum energy in a
Lorentz-breaking scalar field theory is investigated. Also, in Ref.~\cite%
{Erdas:2021xvv}, the author investigates the finite temperature Casimir-like
effect due to a massive and charged scalar field under the influence of a
constant magnetic field, in a Lorentz symmetry violation scenario. A
Casimir-like effect and the generation of topological mass for a system of a real
field interacting with a complex one has recently been considered in Ref.~%
\cite{PhysRevD.107.125019}.

In the present paper we consider two scalar quantum fields, one real and the
other complex, interacting via a quartic interaction and in addition, the
self-interaction of each field is also present. The real field is subject to
a periodic condition, while the components of the complex field are
subjected to a quasi-periodic condition \cite{PhysRevD.107.125019}. The
system is placed in a scenario where the CPT-even aether-type Lorentz
symmetry violation takes place \cite{Car, Gomes:2009ch, Chatrabhuti:2009ew,
Aj}. We allow that the violation occurs with different intensities on each
field. The quasi-periodic condition plays an important role when one
considers nanotubes or nanoloops for a quantum field \cite{feng2014casimir},
for instance, if the phase angle is zero, it corresponds to metallic
nanotubes, while the value of $\pm \frac{2\pi }{3}$ corresponds to a
semiconductor nanotubes. In order to investigate the vacuum energy density
for the mentioned system, we use the path integral formalism and construct
the effective potential in terms of a loop expansion. This formalism was
developed by Jackiw \cite{jackiw1974functional} and allows us to obtain the
vacuum energy density and its loop corrections and in addition, loop
corrections to the mass of the field. The conditions on the fields are
chosen so that we can focus attention on investigating the effect of all
elements present in the system on the vacuum state of the real field. Within
the path integral formalism, we will see that the only constant background
complex field, after the loop expansion, compatible with the imposition of a
quasi-periodic condition is the one that vanishes from the start, something
that does not occur for the real field.

Before proceeding, let us point out that in Ref.~\cite{Jaffe:2005vp} the author argues that the origin of the 
Casimir effect relies on the interaction of the electromagnetic field 
with the physical properties of the plates, i.e., charges and currents. Hence,
it is shown that the original Casimir result for the vacuum force can
be obtained in a framework where it is not necessary to invoke the notion
of quantum vacuum usually present in the standard quantum field theory approach. Although 
in the present work we dot not consider an interaction with possible physical properties
associated with the conditions adopted, we think that it is always important to
describe the Casimir's original configuration as a motivation to pursue an
investigation in other scenarios where modifications on the vacuum 
fluctuations of a quantum field takes place as a consequence of a 
finite length (volume), as in our case. Thus, in contrast with Ref. \cite{Jaffe:2005vp}, 
we make use of the standard quantum field theory framework where the concept
of quantum vacuum is present, and it is widely accepted, to investigate the system described in the 
previous paragraph. However, in order to avoid confusion, instead of using 
terminologies such as `Casimir energy density', or `Casimir force', we shall use
the terminologies `vacuum energy density' or `vacuum force'.

This paper is organized as follows: in Sec.\ref{sec2} we review the main
aspects of the path integral formalism to obtain the effective potential in
the case of two interacting quantum fields, one real and the other complex.
The interaction considered is the so called quartic interaction, that is, a
product between the modulus square of the complex field and the square of
the real field and, in addition, we also consider self-interaction
for each field. The whole system is considered in a scenario where the
CPT-even aether-type Lorentz symmetry violation takes place. In Sec.\ref%
{sec3}, we consider the real and complex fields interacting with each other.
The real field subject to the periodic condition, while the components of
the complex field obey quasi-periodic condition. We assume that the Lorentz
symmetry violation occurs in the $z$-direction, along which the conditions
are imposed and the effects are stronger. Upon using the generalized zeta function
technique, we obtain the effective potential of the system, which allows us
to calculate the vacuum energy density, up to two-loop corrections, and the
one-loop correction to the mass. It is also discussed the conditions for the
stability of the possible vacuum states which guarantees a positive
topological mass. In the end of this section we briefly discuss the effect
of the Lorentz violation scenario considered in the $\tau $, $x$ and $y$
directions. In Sec.\ref{sec5} we present our conclusions. Throughout this
paper we use natural units in which both the Planck constant and the speed
of light are set equal to unit, that is, $\hslash =c=1$, and the metric
signature is $\left( -,+,+,+\right) $.

\section{Effective potential for real and complex scalar fields with Lorentz
violation}

\label{sec2}Here we consider a system composed of a real scalar field
denoted by $\psi $, interacting with a complex field denoted by $\phi $. The
interaction between the fields is represented by a term proportional to the
product of the square of the real field and the modulus square of the
complex field. This choice of interaction is justified since it preservers
the discrete symmetry $\psi \rightarrow -\psi $ of the real field, and also
the global symmetry $\phi \rightarrow e^{i\Theta }\phi $ of the complex
field, which is important for the predicability of a given model \cite{Haber}%
. We also include the self-interaction of each field, which is represented
by terms proportional to the fourth power of each field. The system is
placed in a scenario where the CPT-even aether-type Lorentz symmetry
violation is allowed \cite{Car, Gomes:2009ch, Chatrabhuti:2009ew, Aj}.
Moreover, we work with the Euclidean spacetime coordinates, with imaginary
time \cite{greiner2013field}.

The action describing the system mentioned above, in Euclidean coordinates,
can be written as the following,%
\begin{equation}
S_{\text{E}}\left[ \psi ,\varphi _{i}\right] =S_{\text{K}}\left[ \psi
,\varphi _{i}\right] +S_{\text{L-V}}\left[ \psi ,\varphi _{i}\right] +S_{%
\text{V}}\left[ \psi ,\varphi _{i}\right] ,  \label{rc2}
\end{equation}%
where $S_{\text{K}}\left[ \psi \right] $ is the kinetic term with explicit
form given by \cite{PhysRevD.107.125019},%
\begin{equation}
S_{\text{K}}\left[ \psi ,\varphi _{i}\right] =\frac{1}{2}\int d^{4}x\left(
\psi \square \psi +\sum_{i=1}^{2}\varphi _{i}\square \varphi _{i}\right) .
\end{equation}%
Note that the complex field $\phi $ has been separeted into its two real
components $\varphi _{i}$, $i=1,2$ \cite{ryder1996quantum,
PhysRevD.107.125019}. The D'Alembertian operator, $\square $, is written in
Euclidean spacetime coordinates as%
\begin{equation}
\square =\left( \partial _{\tau }^{2}+\mathbf{\nabla }^{2}\right) .
\end{equation}%
The next term in Eq.~(\ref{rc2}) is the CPT-even aether-type Lorentz
symmetry violation action which is given by \cite{Car, Gomes:2009ch,
Chatrabhuti:2009ew, Aj}%
\begin{equation}
S_{\text{L-V}}\left[ \psi ,\varphi _{i}\right] =\frac{1}{2}\int
d^{4}x\left\{ -\chi _{\psi }\left( u^{0}\partial _{0}\psi \right) ^{2}+\chi
_{\psi }\left( u^{i}\partial _{i}\psi \right) ^{2}+\sum_{i=1}^{2}\left[
-\chi _{\varphi }\left( u^{0}\partial _{0}\varphi _{i}\right) ^{2}+\chi
_{\varphi }\left( u^{j}\partial _{j}\varphi _{i}\right) ^{2}\right] \right\}
.
\end{equation}%
Note that in the above expression, $\chi _{\psi }$ and $\chi _{\varphi }$
characterize the Lorentz violation for each field, and they are supossed to
be much smaller than unity. This way we allow the Lorentz violation to
influence each field with different intensities. The unit vector $u^{\mu }$
determine the direction in which the violation occurs.

%In this way, we
%allowed different intensities for the Lorentz violation for each field, but
%the direction determined by $u^{\mu }$ is the same for the real and complex
%fields. toms1980symmetry, porfirio2021ground

The last term composing the action of the system is the one which includes
the self-interaction and the interaction between the fields, i.e.,%
\begin{eqnarray}
&&S_{\text{V}}\left[ \psi ,\varphi _{i}\right] =-\int d^{4}xU\left( \psi
,\varphi \right) ,  \notag \\
&&U\left( \psi ,\varphi \right) =\frac{\lambda _{\psi }+C_{1}}{4!}\psi ^{4}+%
\frac{m^{2}+C_{2}}{2}\psi ^{2}+\frac{\mu ^{2}}{2}\varphi ^{2}+\frac{g}{2}%
\varphi ^{2}\psi ^{2}+\frac{\lambda _{\varphi }}{4!}\varphi ^{4}+C_{3}.
\end{eqnarray}%
The masses of the real and the complex fields are denoted by $m$ and $\mu $,
respectively, $\lambda _{\psi }$ and $\lambda _{\phi }$ are the constant
couplings of self-interaction and $g$ is the coupling constant of the
interaction between the fields. We also make use of the notation $\varphi
^{2}=\varphi _{1}^{2}+\varphi _{2}^{2}$. The constants $C_{i}$ are the
renormalization constants and its explicit form will be obtained in the
renormalization process of the effective potential. Note that if we set $%
\chi _{\psi }$ and $\chi _{\varphi }$ to zero, we recover the action for the
interacting fields written in \cite{PhysRevD.107.125019}.

The construction of the effective potential using the path integral approach
is described in detail in Refs.~\cite{jackiw1974functional,
ryder1996quantum,toms1980symmetry} (see also \cite%
{cruz2020casimir,porfirio2021ground, PhysRevD.107.125019}). Here we present
only the main steps necessary to our goals. The action written in Eq.~(\ref%
{rc2}), is now expanded about a fixed background $\Psi $, $\Phi _{i}$ that
is, $\psi =\Psi +\chi $, $\varphi _{i}=\Phi _{i}+\varrho $, with $\chi $ and 
$\varrho $ representing quantum fluctuations. Since we are interested in the
real field $\psi $, we seek the effective potential as a function only of $%
\Psi $, i.e., $V_{\mathrm{eff}}\left( \Psi \right) $. Then it is unnecessary
to shift the components of the complex field $\phi $, this amounts to set $%
\Phi _{i}=0$ \cite{jackiw1974functional}. This is because the only fixed
background field $\Phi _{i}$ that can satisfy a quasi-periodic condition, as
in Eq. \eqref{rc16.1}, is the one that vanishes. So we are anticipating this
fact. Note that, as a consequence, we do not need to include counter
terms proportional to the powers of $\phi $. This matter is also treated in
Ref.~\cite{PhysRevD.107.125019}.

The expansion of the effective potential in powers of $\hslash $, up to
order $\hslash ^{2}$, can be written as,%
\begin{equation}
V_{\mathrm{eff}}\left( \Psi \right) =V^{\left( 0\right) }\left( \Psi \right)
+V^{\left( 1\right) }\left( \Psi \right) +V^{\left( 2\right) }\left( \Psi
\right) .  \label{rc2.1}
\end{equation}%
The zero order term, $V^{\left( 0\right) }\left( \Psi \right) $, describes
the classical potential, i.e., the tree-level contribution,%
\begin{equation}
V^{\left( 0\right) }\left( \Psi \right) =U\left( \Psi \right) =\frac{\lambda
_{\psi }+C_{1}}{4!}\Psi ^{4}+\frac{m^{2}+C_{2}}{2}\Psi ^{2}+C_{3},
\label{rc2.2}
\end{equation}%
while $V^{\left( 1\right) }\left( \Psi \right) $ is the one-loop correction
to the classical potential above and is written in terms of a path integral
as \cite{toms1980interacting, PhysRevD.107.125019},%
\begin{eqnarray}
&&V^{\left( 1\right) }\left( \Psi \right) =-\frac{1}{\Omega _{4}}\ln \int 
\mathcal{D}\psi \mathcal{D}\varphi _{1}\mathcal{D}\varphi _{2}\exp \left\{ -%
\frac{1}{2}\left[ \left( \psi ,\hat{A}\psi \right) +\left( \varphi _{1},\hat{%
B}\varphi _{1}\right) +\left( \varphi _{2},\hat{B}\varphi _{2}\right) \right]
\right\} ,  \notag \\
&&\left( \psi ,\hat{A}\psi \right) =\int d^{4}x\ \psi \left( x\right) \hat{A}%
\psi \left( x\right) .  \label{rc2.3}
\end{eqnarray}%
The quantity $\Omega _{4}$ is the 4-dimensional volume of the Euclidean
spacetime, which depends on the conditions imposed on the fields. Note that
since the fixed background field is $\Phi _{i}=0$, there are no crossing terms
in the above exponential \cite{PhysRevD.107.125019}. The self-adjoint
operators $\hat{A}$ and $\hat{B}$ are defined as,%
\begin{eqnarray}
&&\hat{A}=-\square -\chi _{\psi }u^{0}u^{0}\partial _{0}\partial _{0}+\chi
_{\psi }u^{i}u^{j}\partial _{i}\partial _{j}+m^{2}+\frac{\lambda _{\psi }}{2}%
\Psi ^{2},  \notag \\
&&\hat{B}=-\square -\chi _{\varphi }u^{0}u^{0}\partial _{0}\partial
_{0}+\chi _{\varphi }u^{i}u^{j}\partial _{i}\partial _{j}+\mu ^{2}+g\Psi
^{2}.  \label{rc2.5}
\end{eqnarray}

The one-loop correction to the effective potential can be written in terms
of the generalized zeta function constructed from the eigenvalues of the
operators $\hat{A}$ and $\hat{B}$ \cite{toms1980interacting,
PhysRevD.107.125019}. Denoting by $\alpha _{\sigma }$ and $\beta _{\rho }$,
the eigenvalues of the operators $\hat{A}$ and $\hat{B}$, respectively, we
can construct the generalized zeta functions in the following manner,%
\begin{equation}
\zeta _{\alpha }\left( s\right) =\sum_{\sigma }\alpha _{\sigma }^{-s},\
\qquad\qquad\qquad \zeta _{\beta }\left( s\right) =\sum_{\rho }\beta _{\rho
}^{-s},  \label{rc2.6}
\end{equation}%
where $\sigma $ and $\rho $ denote sets of quantum numbers associated with
the eigenfunctions of the operators $\hat{A}$ and $\hat{B}$, respectively.
The summation symbol stands for sum or integration of the quantum numbers,
depending on whether they are discrete or continuous. Once we construct the
generalized zeta functions in Eq.~(\ref{rc2.6}), we write the one-loop
correction to the effective potential as below \cite%
{toms1980symmetry, PhysRevD.107.125019, hawking1977zeta},

\begin{eqnarray}
&&V^{\left( 1\right) }\left( \Psi \right) =V_{\alpha }^{\left( 1\right)
}\left( \Psi \right) +V_{\beta }^{\left( 1\right) }\left( \Psi \right) , 
\notag \\
&&V_{\alpha }^{\left( 1\right) }=-\frac{1}{2\Omega _{4}}\left[ \zeta
_{\alpha }^{\prime }\left( 0\right) +\zeta _{\alpha }\left( 0\right) \ln \nu
^{2}\right] ,  \notag \\
&&V_{\beta }^{\left( 1\right) }=-\frac{1}{\Omega _{4}}\left[ \zeta _{\beta
}^{\prime }\left( 0\right) +\zeta _{\beta }\left( 0\right) \ln \nu ^{2}%
\right] .  \label{rc14}
\end{eqnarray}%
The notation $\zeta _{\alpha ,\beta }\left( 0\right) $ and $\zeta _{\alpha
,\beta }^{\prime }\left( 0\right) $ stands for the generalized zeta function
and its derivative with respect to $s$ evaluated at zero, respectively. The
parameter $\nu $ has dimension of mass and it is an integration measure in
the functional space, which will be removed via renormalization of the
effective potential \cite{toms1980symmetry, toms1980interacting,
PhysRevD.107.125019}. It is convenient to calculate the two-loop correction, 
$V^{\left( 2\right) }\left( \Psi \right) $, of the effective potential from
the two-loop Feynman graph. This correction can also be written it in terms
of the generalized zeta function if one is interested in calculating only
the vacuum contribution \cite{cruz2020casimir,porfirio2021ground}. We
postpone the explicit form of the $V^{\left( 2\right) }\left( \Psi \right) $
until Section \ref{sec3.3}.

Once we obtain the effective potential and its corrections, we have to
perform the renormalization. The renormalization process consists in the
application of the following set of conditions:%
\begin{eqnarray}
&&\left. \frac{d^{4}V_{\mathrm{eff}}\left( \Psi \right) }{d\Psi ^{4}}%
\right\vert _{\Psi =M}=\lambda _{\psi },  \label{rc14.2} \\
&&\left. \frac{d^{2}V_{\mathrm{eff}}\left( \Psi \right) }{d\Psi ^{2}}%
\right\vert _{\Psi =v}=m^{2},  \label{rc14.3} \\
&&\left. V_{\mathrm{eff}}\left( \Psi \right) \right\vert _{\Psi =0}=0.
\label{rc14.5}
\end{eqnarray}%
The first renormalization condition, Eq.~(\ref{rc14.2}), is written in
analogy to Coleman-Weinberg \cite{coleman1973radiative}, fixing the constant 
$C_{1}$ in Eq.~(\ref{rc2}) and also the coupling constant $\lambda _{\psi }$%
. $M$ is a parameter with dimension of mass and can be taken to be zero if
the theory involves only massive fields \cite{cruz2020casimir,
porfirio2021ground, Aj, PhysRevD.107.125019}. The renormalization constant $%
C_3$ is fixed by the condition \eqref{rc14.5}, while the condition in Eq.~(%
\ref{rc14.3}) fix the constant $C_{2}$ in Eq.~(\ref{rc2}), and $\Psi =v$ is
the value of the field that minimizes the potential as long as the extremum
condition is obeyed, that is, 
\begin{equation}
\left. \frac{dV_{\mathrm{eff}}\left( \Psi \right) }{d\Psi }\right\vert
_{\Psi =v}=0.  \label{rc14.8}
\end{equation}%
In the Sec. \ref{sec3.1} we discuss the vacuum stability and present the
values of the field which satisfy the condition above. The last condition,
Eq.~(\ref{rc14.5}), is used only in a massive field theory \cite%
{cruz2020casimir}. The renormalization conditions presented in Eqs.~(\ref%
{rc14.2}), (\ref{rc14.3}) and (\ref{rc14.5}) are to be taken in the limit of
Minkowski spacetime with no conditions imposed.

Now we are well equiped to investigate the loop-expansion of the effective
potential, the generation of topological mass and the vacuum stability for
the system of a real field interacting with a complex field. We will impose
periodic condition for the real field, and quasi-periodic condition for the
components of the complex field as well as consider a Lorentz symmetry violation
for each field.

\section{Periodic and quasi-periodic conditions with Lorentz
violation}

\label{sec3} The action of the system considered here has been written in
Eq.~(\ref{rc2}). The real field is subjected to a periodic condition, while
the components of the complex field obey a quasi-periodic condition. In
addition, we assume that the Lorentz symmetry violation occur in the $z$-direction, which amounts to set the unit vector
as $u^{\mu }=\left( 0,0,0,1\right) $, the same
direction along which the conditions are imposed. In this configuration, the Lorentz violation is expected to be nontrivial in comparison with
the other directions \cite{cruz2020casimir, Aj}. 

Since the real field is subject to the periodic condition together with the
Lorentz symmetry violation in the $z$-direction, the eigenvalues of the
operator $\hat{A}$, presented in Eq.~(\ref{rc2.5}), are found to be \cite%
{Aj, PhysRevD.107.125019},%
\begin{eqnarray}
&&\alpha _{\sigma }=k_{\tau }^{2}+k_{x}^{2}+k_{y}^{2}+\left( 1-\chi _{\psi
}\right) \left( \frac{2\pi n}{L}\right) ^{2}+M_{\lambda }^{2},  \notag \\
&&M_{\lambda }^{2}=m^{2}+\frac{\lambda _{\psi }}{2}\Psi ^{2},\ \ n=0,\pm
1,\pm 2,...,  \label{rc14.1}
\end{eqnarray}
where $L$ is the periodic parameter, which compactifies the $z$-coordinate
into a length $L$. The subscript $\sigma $ stands for the set of quantum
numbers $\left( k_{\tau },k_{x},k_{y},n\right) $.

The components of the complex field are subject to a quasi-periodic
condition as well as to the Lorentz symmetry violation \cite%
{feng2014casimir, porfirio2021ground, Aj, PhysRevD.107.125019}, which leads
to the following set of eigenvalues of the operator $\hat{B}$, presented in
Eq.~(\ref{rc2.5}),%
\begin{eqnarray}
\beta _{\rho } &=&p_{\tau }^{2}+p_{x}^{2}+p_{y}^{2}+\left( 1-\chi _{\varphi
}\right) \left( \frac{2\pi }{L}\right) ^{2}\left( n+\eta \right)
^{2}+M_{g}^{2},  \notag \\
M_{g}^{2} &=&\mu ^{2}+g\Psi ^{2},\ \ n=0,\pm 1,\pm 2,....  \label{rc16.1}
\end{eqnarray}%
Here $\rho $ stands for the set of quantum numbers $\left( p_{\tau
},p_{x},p_{y},n\right) $. Moreover, the allowed values of the phase $\eta $
are in the interval $0\leq \eta <1$. Note that if we set $\eta =0$, then the
field components $\varphi _{i}$ are called untwisted scalar fields and, if $%
\eta =1/2$ we are in the twisted field case \cite{toms1980interacting}.

Before proceeding, we should emphasize here that we wish to consider vacuum 
fluctuation effects on the real scalar field so that the effective potential shall 
depend only on the constant background field $\Psi$, as mentioned in the paragraph
above Eq. \eqref{rc2.1}. This makes possible for the real scalar field to obey
only boundary conditions which are compatible with a nonzero constant $\Psi$. As
pointed out in Ref. \cite{toms1980symmetry, toms1980interacting}, conditions such as antiperiodicity 
which makes the real field to become twisted is only compatible with $\Psi=0$. This,
of course, would make our investigation somewhat restricted. In this sense, we have adopted
periodic boundary condition for the real scalar field in order to circumvent this problem and make our analysis as general as possible. On the other hand, as it is intended 
in the present work, we can consider for instance an antiperiodic complex scalar field ($\eta=1/2$ in Eq. \eqref{rc16.1}) coupled to the real scalar field,
something very similar to what has been done in Ref. \cite{toms1980interacting} (see also Refs. \cite{Ford:1979pr, Oikonomou:2006ar, deFarias:2020xms} for twisted scalar fields considered in other contexts). Our goal will then be achieved 
once we calculate the effects of this complex scalar field on the effective potential as a function of $\Psi$, associated with the real scalar field.
This, as a consequence, will make possible to obtain the vacuum energy density and generation of topological mass in a consistent manner.

Once we know the explicit form of the eigenvalues $\alpha _{\sigma }$ and $%
\beta _{\rho }$, given in Eqs. (\ref{rc14.1}) and (\ref{rc16.1}),
respectively, we construct the generalized zeta function, Eq.~(\ref{rc2.6}),
and obtain a useful expression for the first order correction to the
effetive potential, Eq.~(\ref{rc14}). The vacuum energy density and also the
topological mass are obtained from the renormalized effective potential
which in turn is obtained from effective potential after the application of
the renormalization conditions. In the next section we shall obtain the
renormalized effective potential, up to first order correction, and the
vacuum energy density.

\subsection{One-loop correction and the vacuum energy density}

We start considering first the one-loop correction from the real field $\psi 
$. The eigenvalues used to construct the generalized zeta function, Eq.~(\ref%
{rc2.6}), are presented in Eq.~(\ref{rc14.1}). Therefore, the generalized
zeta function takes the following form:%
\begin{equation}
\zeta _{\alpha z}\left( s\right) =\frac{\Omega _{3}}{\left( 2\pi \right) ^{3}%
}\int dk_{\tau }dk_{x}dk_{y}\sum_{n=-\infty }^{+\infty }\left\{ k_{\tau
}^{2}+k_{x}^{2}+k_{y}^{2}+\left( 1-\chi _{\psi }\right) \left( \frac{2\pi n}{%
L}\right) ^{2}+M_{\lambda }^{2}\right\} ^{-s},  \label{rc1.1}
\end{equation}%
where $\Omega _{3}$ represents the 3-volume associated with the Euclidean
spacetime coordinates $\tau ,x,z$, necessary to make the integrals
dimensionless. The subscript $z$ is to remind that we are in the case where
the Lorentz violation occur in the $z$-direction. In order to obtain a
practical expression for the generalized zeta function (\ref{rc1.1}), we
take similar steps as the ones presented in \cite{porfirio2021ground,
PhysRevD.107.125019}, hence, we shall indicate the main steps. The first
one is to use the following identity,%
\begin{equation}
w^{-s}=\frac{2}{\Gamma \left( s\right) }\int_{0}^{\infty }d\tau \ \tau
^{2s-1}e^{-w\tau ^{2}},  \label{i1}
\end{equation}%
and then perform the resulting Gaussians integrals in $k_{\tau }$, $k_{x}$
and $k_{y}$. After that, we identify the integral representation of the
gamma function $\Gamma \left( z\right) $ \cite{abramowitz1965handbook},%
\begin{equation}
\frac{\Gamma \left( z\right) }{2}=\int_{0}^{\infty }d\mu \ \mu
^{2z-1}e^{-\mu ^{2}},  \label{intg}
\end{equation}%
yielding the following form of the generalized zeta function,%
\begin{equation}
\zeta _{\alpha z}\left( s\right) =\frac{\Omega _{4}\pi ^{\frac{3}{2}-2s}}{%
2^{2s}L^{4-2s}}\frac{\Gamma \left( s-\frac{3}{2}\right) }{\Gamma \left(
s\right) }\left( 1-\chi _{\psi }\right) ^{\frac{3}{2}-s}\sum_{n=-\infty
}^{+\infty }\left[ n^{2}+\left( \frac{M_{\lambda }L}{2\pi \sqrt{1-\chi
_{\psi }}}\right) ^{2}\right] ^{\frac{3}{2}-s}.  \label{z2}
\end{equation}%
In the above expression, $\Omega _{4}$ is the 4-dimensional volume in
Euclidean spacetime, which takes into account the spacetime topology. In the
case under consideration, the 4-dimensional volume is written as $\Omega
_{4}=\Omega _{3}L$. The sum in Eq.~(\ref{z2}) is performed via the folowing
analytic continuation of the inhomogeneous, generalized Epstein function 
\cite{elizalde1995zeta,feng2014casimir, PhysRevD.107.125019}:

\begin{equation}
\sum_{n=-\infty }^{+\infty }\left[ \left( n+\vartheta \right) ^{2}+\kappa
^{2}\right] ^{-z}=\frac{\pi ^{\frac{1}{2}}\kappa ^{1-2z}}{\Gamma \left(
z\right) }\left\{ \Gamma \left( z-\frac{1}{2}\right) +4\left( \pi \kappa
\right) ^{z-\frac{1}{2}}\sum_{j=1}^{\infty }j^{z-\frac{1}{2}}\cos \left(
2\pi j\vartheta \right) K_{\left( \frac{1}{2}-z\right) }\left( 2\pi j\kappa
\right) \right\} ,  \label{id2}
\end{equation}%
where $K_{\gamma }(x)$ is the modified Bessel function of the second kind
or, as it is also known, the Macdonald function \cite{abramowitz1965handbook}%
. With the help of Eq.~(\ref{id2}), one obtains the generalized zeta
function, Eq.~(\ref{z2}), in a practical form, i.e.,%
\begin{equation}
\zeta _{\alpha z}\left( s\right) =\frac{\Omega _{4}}{2^{4}\pi ^{2}}\frac{%
M_{\lambda }^{4-2s}}{\sqrt{1-\chi _{\psi }}\Gamma \left( s\right) }\left\{
\Gamma \left( s-2\right) +2^{4-s}\sum_{j=1}^{\infty }f_{\left( 2-s\right)
}\left( \frac{jM_{\lambda }L}{\sqrt{1-\chi _{\psi }}}\right) \right\} ,
\label{rc1.2}
\end{equation}%
where we have defined the function $f_{\gamma }\left( x\right) $ as the ratio%
\begin{equation}
f_{\gamma }\left( x\right) =\frac{K_{\gamma }\left( x\right) }{x^{\gamma }}.
\label{rc1.4}
\end{equation}

Therefore, the one-loop correction $V_{\alpha z}^{\left( 1\right) }\left(
\Psi \right) $, Eq.~(\ref{rc14}), to the effective potential associated with
the real field is written in the form%
\begin{equation}
V_{\alpha z}^{\left( 1\right) }\left( \Psi \right) =\frac{M_{\lambda }^{4}}{%
2^{6}\pi ^{2}\sqrt{1-\chi _{\psi }}}\left[ \ln \left( \frac{M_{\lambda }^{2}%
}{\nu ^{2}}\right) -\frac{3}{2}\right] -\frac{M_{\lambda }^{4}}{2\pi ^{2}%
\sqrt{1-\chi _{\psi }}}\sum_{j=1}^{\infty }f_{2}\left( \frac{jM_{\lambda }L}{%
\sqrt{1-\chi _{\psi }}}\right) .  \label{rc1.3}
\end{equation}%
The expression above is the first order correction, that is, the one-loop
correction to the effective potential due to the real field subject to the
periodic condition, with Lorentz symmetry violation in the $z$-direction.
Now we have to calculate the contribution due to the complex field.

The eigenvalues of the operator $\hat{B}$, associated with the complex field
components are presented in Eq.~(\ref{rc16.1}) and leads to the following
generalized zeta function,%
\begin{equation}
\zeta _{\beta z}\left( s\right) =\frac{\Omega _{3}}{\left( 2\pi \right) ^{3}}%
\int dp_{\tau }dp_{x}dp_{y}\sum_{n=-\infty }^{+\infty }\left[ p_{\tau
}^{2}+p_{x}^{2}+p_{y}^{2}+\left( 1-\chi _{\varphi }\right) \left( \frac{2\pi 
}{L}\right) ^{2}\left( n+\eta \right) ^{2}+M_{g}^{2}\right] ^{-s}.
\label{rc1.5}
\end{equation}%
A simplified expression for $\zeta _{\beta z}\left( s\right) $ can be obtained
in a similar way as the previous one \cite{porfirio2021ground,
PhysRevD.107.125019}. Therefore we present the final form of the generalized
zeta function $\zeta _{\beta z}\left( s\right) $,%
\begin{equation}
\zeta _{\beta z}\left( s\right) =\frac{\Omega _{4}}{2^{4}\pi ^{2}}\frac{%
M_{g}^{4-2s}}{\sqrt{1-\chi _{\varphi }}\Gamma \left( s\right) }\left\{
\Gamma \left( s-2\right) +2^{4-s}\sum_{j=1}^{\infty }\cos \left( 2\pi j\eta
\right) f_{\left( 2-s\right) }\left( \frac{jM_{g}L}{\sqrt{1-\chi _{\varphi }}%
}\right) \right\} .  \label{rc1.6}
\end{equation}%
Calculating the zeta function above and its derivative at $s=0$, we find the
one-loop correction to the effective potential, Eq.~(\ref{rc14}), due to the
complex field as 
\begin{equation}
V_{\beta z}^{\left( 1\right) }\left( \Psi \right) =\frac{M_{g}^{4}}{2^{5}\pi
^{2}\sqrt{1-\chi _{\varphi }}}\left[ \ln \left( \frac{M_{g}^{2}}{\nu ^{2}}%
\right) -\frac{3}{2}\right] -\frac{M_{g}^{4}}{\pi ^{2}\sqrt{1-\chi _{\varphi
}}}\sum_{j=1}^{\infty }\cos \left( 2\pi j\eta \right) f_{\left( 2\right)
}\left( \frac{jM_{g}L}{\sqrt{1-\chi _{\varphi }}}\right) .  \label{rc1.7}
\end{equation}

Collecting the results presented in Eqs.~(\ref{rc1.3}) and (\ref{rc1.7}),
one can write the first order, or the one-loop correction to the effective
potential in the following form,%
\begin{eqnarray}
&&V_{z}^{\left( 1\right) }\left( \Psi \right) =\frac{M_{\lambda }^{4}}{%
2^{6}\pi ^{2}\sqrt{1-\chi _{\psi }}}\left[ \ln \left( \frac{M_{\lambda }^{2}%
}{\nu ^{2}}\right) -\frac{3}{2}\right] +\frac{M_{g}^{4}}{2^{5}\pi ^{2}\sqrt{%
1-\chi _{\varphi }}}\left[ \ln \left( \frac{M_{g}^{2}}{\nu ^{2}}\right) -%
\frac{3}{2}\right]  \notag \\
&&-\frac{M_{\lambda }^{4}}{2\pi ^{2}\sqrt{1-\chi _{\psi }}}%
\sum_{n=1}^{\infty }f_{2}\left( \frac{nM_{\lambda }L}{\sqrt{1-\chi _{\psi }}}%
\right) -\frac{M_{g}^{4}}{\pi ^{2}\sqrt{1-\chi _{\varphi }}}%
\sum_{j=1}^{\infty }\cos \left( 2\pi j\eta \right) f_{2}\left( \frac{jM_{g}L%
}{\sqrt{1-\chi _{\varphi }}}\right) .  \label{rc1.8}
\end{eqnarray}%
Therefore, the nonrenormalized effective potential, Eq.~(\ref{rc14}), up to
order $\hslash $, reads,%
\begin{eqnarray}
&&V_{\mathrm{eff}}\left( \Psi \right) _{z}=\frac{m^{2}+C_{2}}{2}\Psi ^{2}+%
\frac{\lambda _{\psi }+C_{1}}{4!}\Psi ^{4}+C_{3}  \notag \\
&&+\frac{M_{\lambda }^{4}}{2^{6}\pi ^{2}\sqrt{1-\chi _{\psi }}}\left[ \ln
\left( \frac{M_{\lambda }^{2}}{\nu ^{2}}\right) -\frac{3}{2}\right] +\frac{%
M_{g}^{4}}{2^{5}\pi ^{2}\sqrt{1-\chi _{\varphi }}}\left[ \ln \left( \frac{%
M_{g}^{2}}{\nu ^{2}}\right) -\frac{3}{2}\right]  \notag \\
&&-\frac{M_{\lambda }^{4}}{2\pi ^{2}\sqrt{1-\chi _{\psi }}}%
\sum_{n=1}^{\infty }f_{2}\left( \frac{nM_{\lambda }L}{\sqrt{1-\chi _{\psi }}}%
\right) -\frac{M_{g}^{4}}{\pi ^{2}\sqrt{1-\chi _{\varphi }}}%
\sum_{j=1}^{\infty }\cos \left( 2\pi j\eta \right) f_{2}\left( \frac{jM_{g}L%
}{\sqrt{1-\chi _{\varphi }}}\right) .  \label{nrepz}
\end{eqnarray}

The renormalization constants $C_{i}$ are obtained by applying the
renormalization conditions expressed in the Eqs.~(\ref{rc14.2}), (\ref%
{rc14.3}) and (\ref{rc14.5}), in the limit of Minkowski spacetime, $%
L\rightarrow \infty $. The explicit form of the $C_{i}$'s are%
\begin{eqnarray}
&&C_{1}=\frac{6\lambda _{\psi }^{2}}{2^{6}\pi ^{2}\sqrt{1-\chi _{\psi }}}\ln
\left( \frac{\nu ^{2}}{m^{2}}\right) +\frac{24g^{2}}{2^{5}\pi ^{2}\sqrt{%
1-\chi _{\varphi }}}\ln \left( \frac{\nu ^{2}}{\mu ^{2}}\right) ,  \notag \\
&&C_{2}=\frac{2\lambda _{\psi }m^{2}}{2^{6}\pi ^{2}\sqrt{1-\chi _{\psi }}}%
\left[ \ln \left( \frac{\nu ^{2}}{m^{2}}\right) +1\right] +\frac{4g\mu ^{2}}{%
2^{5}\pi ^{2}\sqrt{1-\chi _{\varphi }}}\left[ \ln \left( \frac{\nu ^{2}}{\mu
^{2}}\right) +1\right] ,  \notag \\
&&C_{3}=\frac{m^{4}}{2^{6}\pi ^{2}\sqrt{1-\chi _{\psi }}}\left[ \ln \left( 
\frac{\nu ^{2}}{m^{2}}\right) +\frac{3}{2}\right] +\frac{\mu ^{4}}{2^{5}\pi
^{2}\sqrt{1-\chi _{\varphi }}}\left[ \ln \left( \frac{\nu ^{2}}{\mu ^{2}}%
\right) +\frac{3}{2}\right] .
\end{eqnarray}%
Note that all coefficients above depend on the Lorentz violation parameters $\chi _{\psi }$ and $\chi
_{\varphi}$. Upon substituting the renormalization constants above into the
effective potential, Eq.~(\ref{nrepz}), we obtain the renormalized effective
potential for a system of interacting scalar fields with the Lorentz
violation in the $z$-direction:%
\begin{eqnarray}
&&V_{\mathrm{eff}}^{R}\left( \Psi \right) _{z}=\frac{m^{2}}{2}\Psi ^{2}+%
\frac{\lambda _{\psi }}{4!}\Psi ^{4}+\frac{m^{4}}{2^{6}\pi ^{2}\sqrt{1-\chi
_{\psi }}}\ln \left( \frac{M_{\lambda }^{2}}{m^{2}}\right) +\frac{\mu ^{4}}{%
2^{5}\pi ^{2}\sqrt{1-\chi _{\varphi }}}\ln \left( \frac{M_{g}^{2}}{\mu ^{2}}%
\right)  \notag \\
&&+\frac{\lambda _{\psi }^{2}\Psi ^{4}}{2^{8}\pi ^{2}\sqrt{1-\chi _{\psi }}}%
\left[ \ln \left( \frac{M_{\lambda }^{2}}{m^{2}}\right) -\frac{3}{2}\right] +%
\frac{\lambda _{\psi }m^{2}\Psi ^{2}}{2^{6}\pi ^{2}\sqrt{1-\chi _{\psi }}}%
\left[ \ln \left( \frac{M_{\lambda }^{2}}{m^{2}}\right) -\frac{1}{2}\right] 
\notag \\
&&+\frac{g^{2}\Psi ^{4}}{2^{5}\pi ^{2}\sqrt{1-\chi _{\varphi }}}\left[ \ln
\left( \frac{M_{g}^{2}}{\mu ^{2}}\right) -\frac{3}{2}\right] +\frac{g\mu
^{2}\Psi ^{2}}{2^{4}\pi ^{2}\sqrt{1-\chi _{\varphi }}}\left[ \ln \left( 
\frac{M_{g}^{2}}{\mu ^{2}}\right) -\frac{1}{2}\right]  \notag \\
&&-\frac{M_{\lambda }^{4}}{2\pi ^{2}\sqrt{1-\chi _{\psi }}}%
\sum_{n=1}^{\infty }f_{2}\left( \frac{nM_{\lambda }L}{\sqrt{1-\chi _{\psi }}}%
\right) -\frac{M_{g}^{4}}{\pi ^{2}\sqrt{1-\chi _{\varphi }}}%
\sum_{j=1}^{\infty }\cos \left( 2\pi j\eta \right) f_{2}\left( \frac{jM_{g}L%
}{\sqrt{1-\chi _{\varphi }}}\right) .  \label{rcq4}
\end{eqnarray}

The vacuum energy density is evaluated in a straightforward way once we know
the explicit form of the renormalized effective potential presented in Eq.~(%
\ref{rcq4}). Although there are in fact three possible vacuum, as we will se
in section \ref{sec3.1}, here we consider the vacuum state as $\Psi =0$.
Hence, the vacuum energy density is obtained by taking $\Psi =0$ in the
renormalized effective potential, i.e.,%
\begin{equation}
\mathcal{E}_{0z}=\left. V_{\mathrm{eff}}^{R}\left( \Psi \right)
_{z}\right\vert _{\Psi =0}=-\frac{m^{4}}{2\pi ^{2}\sqrt{1-\chi _{\psi }}}%
\sum_{n=1}^{\infty }f_{2}\left( \frac{nmL}{\sqrt{1-\chi _{\psi }}}\right) -%
\frac{\mu ^{4}}{\pi ^{2}\sqrt{1-\chi _{\varphi }}}\sum_{j=1}^{\infty }\cos
\left( 2\pi j\eta \right) f_{2}\left( \frac{j\mu L}{\sqrt{1-\chi _{\varphi }}%
}\right) .  \label{rc1.15}
\end{equation}%
The first term on the r.h.s of Eq.~(\ref{rc1.15}) is the contribution to the
vacuum energy density from the real field $\psi $ under periodic condition,
while the second term is the contribution from the complex field components,
which is twice the contribution from one Klein-Gordon field under
quasi-periodic condition \cite{porfirio2021ground}. Obviously these two
terms shows the dependence on the Lorentz violation parameters. Recalling
that the Lorentz violation occurs in the $z$-direction, and comparing with
the result obtained in Ref.~\cite{PhysRevD.107.125019} we see the difference
in the multiplicative factor in the form $\left( 1-\chi \right) ^{-\frac{1}{2%
}}$ which is also present in the argument of the Macdonald function, for
each field contribution separately.\newline
\begin{figure}[tbp]
\includegraphics[width=.5\textwidth]{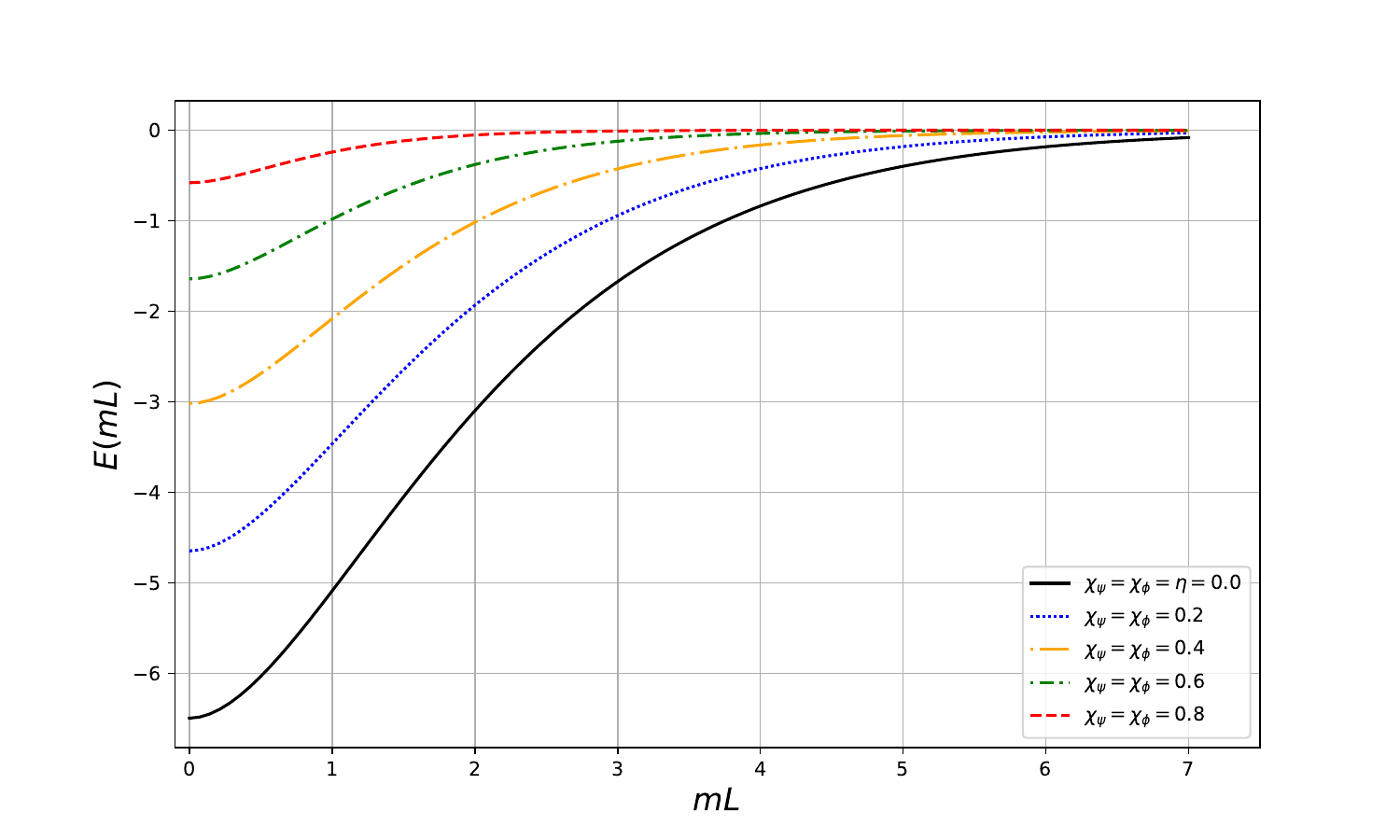}\hfill %
\includegraphics[width=.5\textwidth]{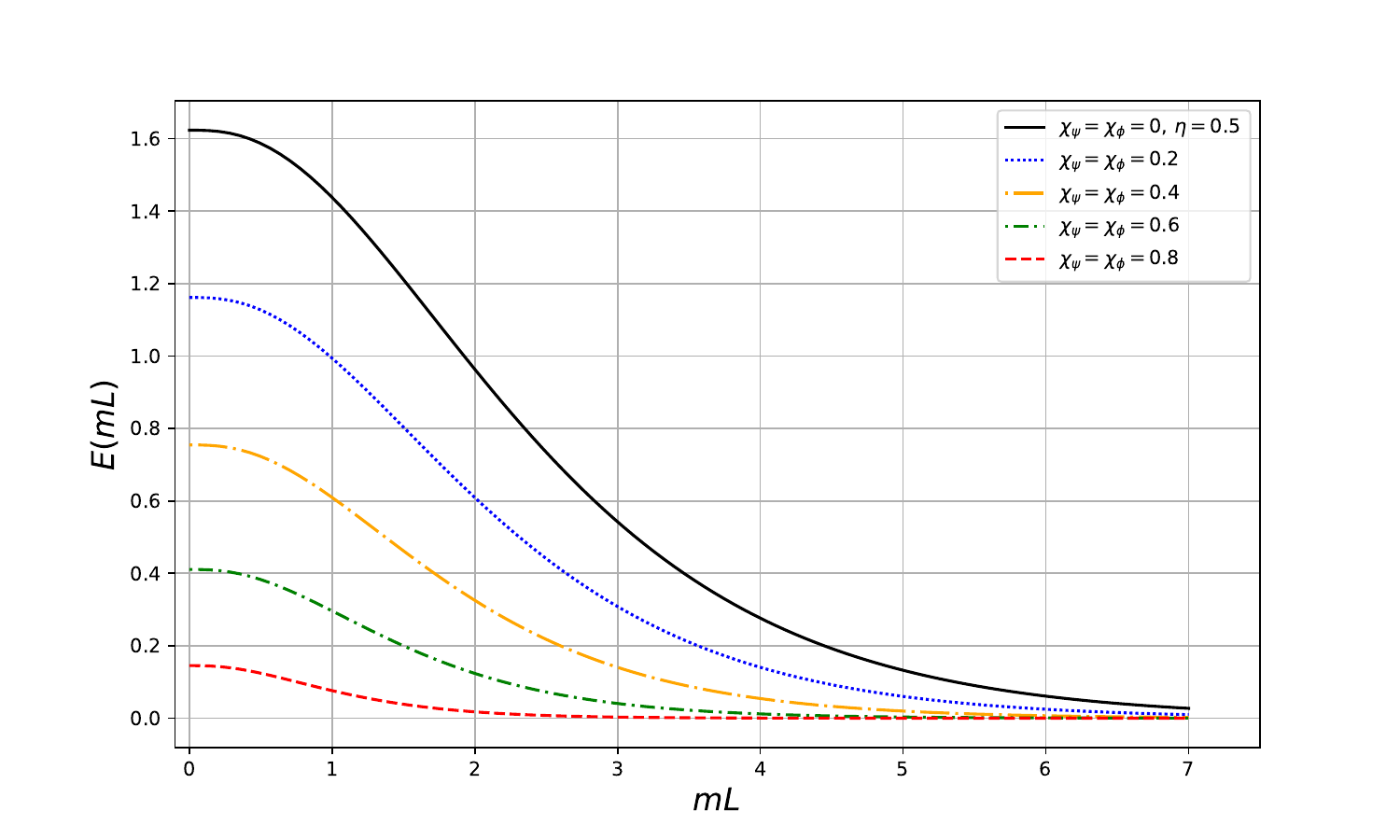}
\caption{Graph of the dimensionless vacuum energy density $E\left(
mL\right) =2\protect\pi ^{2}L^{4}\mathcal{E}_{0z}\left(
mL\right) $, considering $m=\protect\mu $ and $\protect\chi _{\protect\psi }=%
\protect\chi _{\protect\varphi }$.}
\label{fig1}
\end{figure}

In Fig.~\ref{fig1} we exhibit the dimensionless vacuum energy density $%
E\left( mL\right) =2\pi ^{2}L^{4}\mathcal{E}_{0z}\left(
mL\right) $, Eq.~(\ref{rc1.15}), as a function of $mL$, for several values
of the parameters $\chi _{\psi }=\chi _{\varphi }$. We also consider equal
masses, that is, $m=\mu $. The graph on the left shows that for a fixed
value of the quasi-periodic parameter $\eta =0$, as we increase the value of
the parameters $\chi _{\psi }$ and $\chi _{\varphi }$, the vacuum energy
density also increases, comparing its value with the one in which the
Lorentz symmetry is preserved (black solid line). The graph on the right
takes the value $\eta =0.5$, showing that as we increase the values of $\chi
_{\psi }$ and $\chi _{\varphi }$, the vacuum energy density decreases. Note
that the vacuum energy density can be positive or negative, depending on the
value of the quasi-periodic parameter $\eta $. Taking the limit $%
mL\rightarrow \infty $, the vacuum energy density goes to zero, as it
should, since we should not expect any quantum effect in this regime. 
\newline
\begin{figure}[tbp]
\includegraphics[width=.5\textwidth]{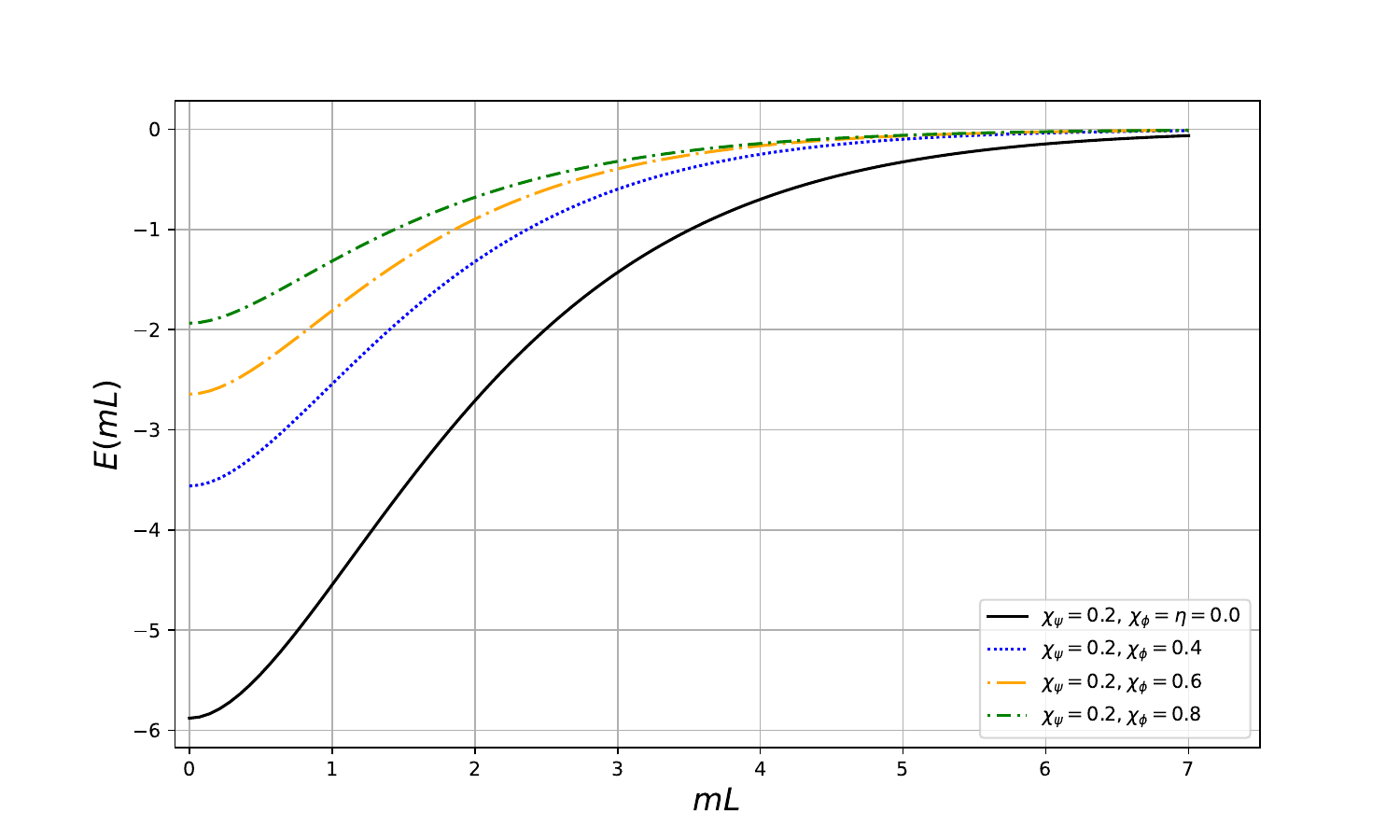}\hfill %
\includegraphics[width=.5\textwidth]{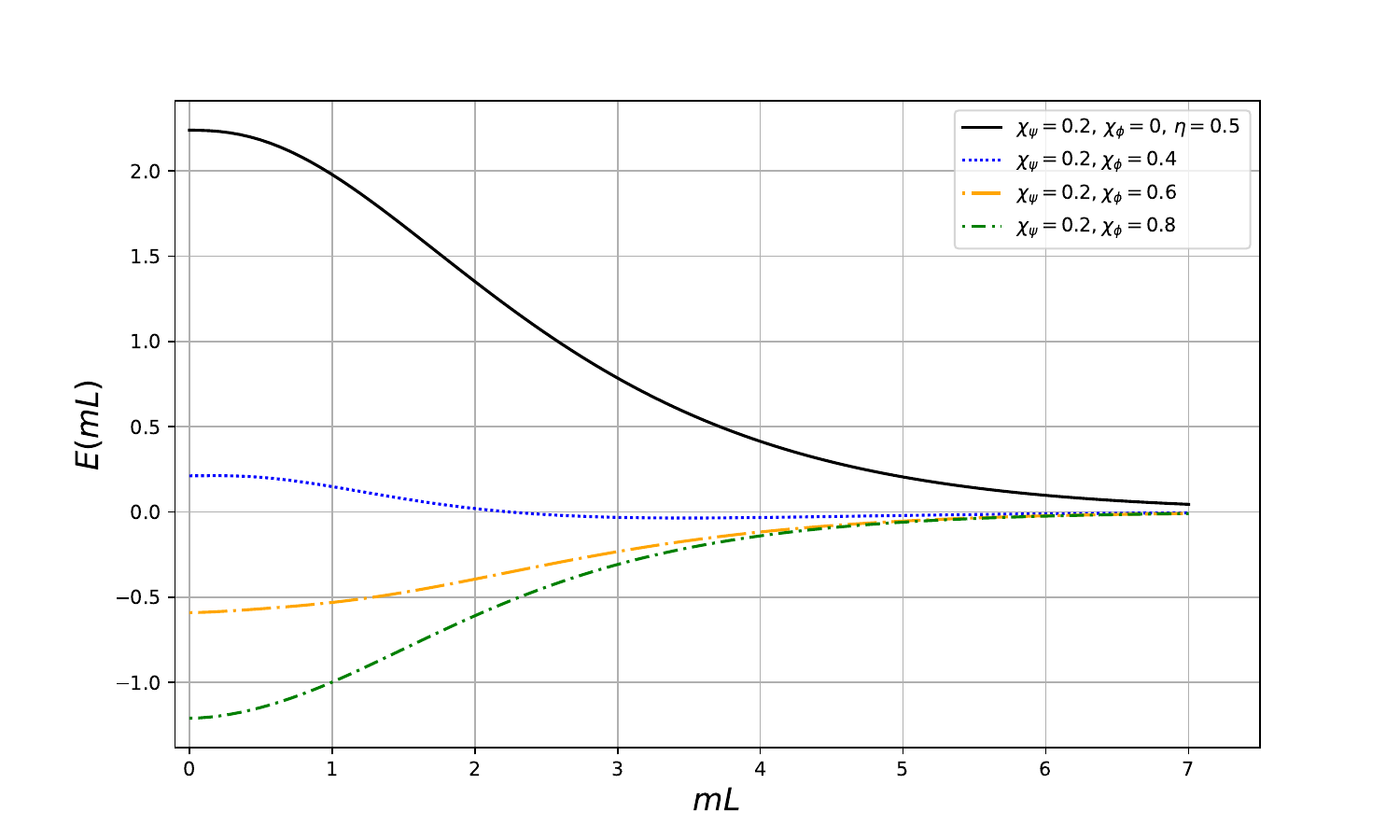}
\caption{Graph of the dimensionless vacuum energy density $E\left(
mL\right) =2\protect\pi ^{2}L^{4}\mathcal{E}_{0z}\left(
mL\right) $, considering $m=\protect\mu $ and $\protect\chi _{\protect\psi %
}=0.2$ with some different values of $\protect\chi _{\protect\varphi }$.}
\label{fig2}
\end{figure}

In Fig.~\ref{fig2} the dimensionless vacuum energy density $E\left(
mL\right) =2\pi ^{2}L^{4}\mathcal{E}_{0z}\left( mL\right) $,
from Eq.~(\ref{rc1.15}), is shown considering the value of the Lorentz
violation parameter for the real field fixed, $\chi _{\psi }=0.2$, and
sevreal values for the Lorentz violation parameter associated with the
complex field $\chi _{\varphi }$. The graph on the left is for $\eta =0$
showing that as we increase the value of $\chi _{\varphi }$, the vacuum
energy density also increases. The graph on the right considers $\eta =0.5$
and shows that as $\chi _{\varphi }$ increases, the vacuum energy density
decreases. In addition, the dotted blue line which takes the values $\chi
_{\psi }=0.2$ and $\chi _{\varphi }=0.4$ exhibits an interesting behaviour,
starting with positive values of $E\left( mL\right) $, becoming negative and
eventually going to zero again. As we can see, the vacuum energy density can
be positive or negative, depending on the values of the parameters of the
Lorentz violation and also, as $mL\rightarrow \infty $, the vacuum energy
density goes to zero.\newline

From the result presented in Eq.~(\ref{rc1.15}), we can consider the
massless fields case in which we take the limit $m,\mu \rightarrow 0$. In
order to obtain the vacuum energy density for the massless fields case, one
can use the limit of small arguments of the Macdonald function, i.e., $%
K_{\mu }\left( x\right) \approx \frac{\Gamma \left( \mu \right) }{2}\left( 
\frac{2}{x}\right) ^{\mu }$ \cite{abramowitz1965handbook}. Hence, in the
exact limit where the masses vanishe, we obtain,%
\begin{equation}
\mathcal{E}_{0z}=-\left( 1-\chi _{\psi }\right) ^{\frac{3}{2}}%
\frac{\pi ^{2}}{90L^{4}}-\frac{2\left( 1-\chi _{\varphi }\right) ^{\frac{3}{2%
}}}{\pi ^{2}L^{4}}\sum_{j=1}^{\infty }j^{-4}\cos \left( 2\pi j\eta \right) ,
\label{rc1.15.a}
\end{equation}%
where we have used the following result for the Riemann zeta function $\zeta
_{\text{R}}\left( 4\right) =\frac{\pi ^{4}}{90}$ \cite%
{elizalde1995zeta,elizalde1995ten}, in the first term on the r.h.s. of Eq.~(%
\ref{rc1.15.a}). This term is an already known result \cite{toms1980symmetry}
multiplied by the factor due to the Lorentz symmetry violation, i.e., $%
\left( 1-\chi _{\psi }\right) ^{\frac{3}{2}}$. The second term can be
rewritten in terms of the Bernoulli polynomials, defined as,%
\begin{equation}
B_{2k}\left( \theta \right) =\frac{\left( -1\right) ^{k-1}2\left( 2k\right) !%
}{\left( 2\pi \right) ^{2k}}\sum_{n=1}^{\infty }\frac{\cos \left( 2\pi
n\theta \right) }{n^{2k}}.  \label{rc1.15.b}
\end{equation}%
Hence, one obtains the following expression for the vacuum energy density,
for the massless field case,%
\begin{equation}
\mathcal{E}_{0z}=-\left( 1-\chi _{\psi }\right) ^{\frac{3}{2}}%
\frac{\pi ^{2}}{90L^{4}}+\left( 1-\chi _{\varphi }\right) ^{\frac{3}{2}}%
\frac{2\pi ^{2}}{3L^{4}}\left( \eta ^{4}-2\eta ^{3}+\eta ^{2}-\frac{1}{30}%
\right) ,  \label{rc1.16}
\end{equation}%
where the Bernoulli polynomial of fourth order, $B_{4}\left( \theta \right)
=\left( \theta ^{4}-2\theta ^{3}+\theta ^{2}-\frac{1}{30}\right) $, has been
used. \ As one should expect, if we take $\chi _{\psi },\chi _{\varphi
}\rightarrow 0$ the first term on the r.h.s. of Eq.~(\ref{rc1.16}) is
consistent with the resulst found in \cite{toms1980symmetry} and, the second
term is consistent with the resulst found in \cite{feng2014casimir} (taking
into accont the two components of the complex field). Therefore the
modification due to the Lorentz symmetry violation is expressed by the
multiplicative factors $\left( 1-\chi _{\psi }\right) ^{\frac{3}{2}}$, for
the contribution due to the real field, and $\left( 1-\chi _{\varphi
}\right) ^{\frac{3}{2}}$ for the contribution from the complex field. Note
that the vacuum energy density can be positive or negative depending on the value
of the parameter $\eta $ of the quasi-periodic condition obeyed by the
components of the complex field.

Now we proceed to calculate the two-loop correction to the effective
potential, in the vacuum state $\Psi =0$, which is the first order $\lambda$
correction to the vacuum energy density.

\subsection{Two-loop correction}

\label{sec3.3}As we have mentioned in the first section of the present
paper, the correction for the vacuum energy density which is first
order in the coupling constants, is obtained from the two-loop Feynman
graphs of the theory. Since we have more than one contribution, we evaluate
the two-loop contributions from each Feynman graph separately. Hence, we
write $V^{\left( 2\right) }\left( \Psi \right) $ as a sum of four terms,%
\begin{equation}
V^{\left( 2\right) }\left( \Psi \right) =V_{\lambda _{\psi }}^{\left(
2\right) }\left( \Psi \right) +V_{\lambda _{\varphi }}^{\left( 2\right)
}\left( \Psi \right) +V_{g}^{\left( 2\right) }\left( \Psi \right)
+V_{2\lambda _{\varphi }}^{\left( 2\right) }\left( \Psi \right) .
\label{2lc}
\end{equation}%
Each term on the r.h.s. of the above equation is associated with a term of
the potential in Eq.~(\ref{rc2}), namely, $V_{\lambda _{\psi }}^{\left(
2\right) }$ is the contribution from the self-interaction of the real field $%
\left( \frac{\lambda _{\psi }}{4!}\psi ^{4}\right) $, $V_{\lambda _{\varphi
}}^{\left( 2\right) }$ is associated with the self-interaction of the
complex field $\left( \frac{\lambda _{\varphi }}{4!}\varphi _{i}^{4}\right) $
, $V_{g}^{\left( 2\right) }$ is associated with the interaction between the
real and complex fields $\left( \frac{g}{2}\varphi _{i}^{2}\psi ^{2}\right) $%
, and finally $V_{2\lambda _{\varphi }}^{\left( 2\right) }$ is associated
with the cross terms arising from the interaction between the components of
the complex field $\left( \frac{\lambda _{\varphi }}{4!}2\varphi
_{1}^{2}\varphi _{2}^{2}\right) $, which is also a self-interaction.\\

\begin{figure}
\centering
\includegraphics[scale=0.8]{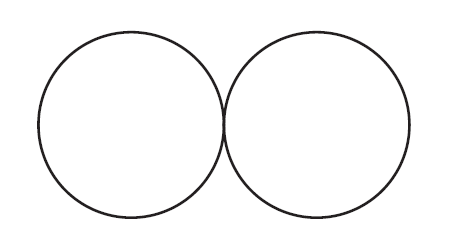}
\caption{Feynman graph representing the contribution from the
self-interaction potential at $\Psi =0$.}
\label{fig3}
\end{figure}

Let us first consider the contribution from the self-interaction term
associated with the real field, that is, $V_{\lambda _{\psi }}^{\left(
2\right) }\left( \Psi \right) $. Since we are interested in the vacuum state
where $\Psi =0$, the only non-vanishing contribution comes from the graph
exhibited in Fig.~\ref{fig3}. From this Feynman graph one can write the
two-loop contribution in terms of the generalized zeta function presented in
Eq.~(\ref{rc1.2}), in the following form \cite%
{Aj,porfirio2021ground,PhysRevD.107.125019},%
\begin{equation}
V_{\lambda _{\psi }}^{\left( 2\right) }\left( 0\right) =\frac{\lambda _{\psi
}}{8}\left. \left[ \frac{\zeta _{\alpha }^{R}\left( 1\right) }{\Omega _{4}}%
\right] ^{2}\right\vert _{\Psi =0}.  \label{rcq7}
\end{equation}%
The zeta function $\zeta _{\alpha }^{R}\left( s\right) $, is defined as the
non-divergent part of the generalized zeta function given in Eq.~(\ref{rc1.2}%
), at $s=1$ \cite{Aj,porfirio2021ground,PhysRevD.107.125019}, i.e.,%
\begin{eqnarray}
\zeta _{\alpha }^{R}\left( s\right) &=&\zeta _{\alpha }\left( s\right) -%
\frac{\Omega _{4}\left( M_{\lambda }\right) ^{4-2s}}{2^{4}\pi ^{2}\sqrt{%
1-\chi _{\psi }}}\frac{\Gamma \left( s-2\right) }{\Gamma \left( s\right) } 
\notag \\
&=&\frac{\left( M_{\lambda }\right) ^{2}}{2\pi ^{2}\sqrt{1-\chi _{\psi }}}%
\sum_{n=1}^{\infty }f_{\left( 1\right) }\left( \frac{nM_{\lambda }L}{\sqrt{%
1-\chi _{\psi }}}\right) .  \label{rcq8}
\end{eqnarray}%
Note that the term that is being subtracted in the above expression is
independent of the parameter $L$ characterizing the conditions and, as usual,
it should be dropped. Explicitly, one obtains the following result for the
two-loop contribution due to the self-interaction term of the real field,%
\begin{equation}
V_{\lambda _{\psi }}^{\left( 2\right) }\left( 0\right) =\frac{\lambda _{\psi
}m^{4}}{32\pi ^{4}\left( 1-\chi _{\psi }\right) }\left[ \sum_{n=1}^{\infty
}f_{\left( 1\right) }\left( \frac{nmL}{\sqrt{\left( 1-\chi _{\psi }\right) }}%
\right) \right] ^{2},  \label{rcq9}
\end{equation}%
which shows that the two-loop contribution is proportional to the coupling
constant $\lambda _{\psi }$ as it should.

Similarly, the other contributions are written as%
\begin{eqnarray}
&&V_{\lambda _{\varphi }}^{\left( 2\right) }\left( 0\right) =2\frac{\lambda
_{\varphi }\mu ^{4}}{32\pi ^{4}\left( 1-\chi _{\varphi }\right) }\left[
\sum_{j=1}^{\infty }\cos \left( 2\pi j\eta \right) f_{\left( 1\right)
}\left( \frac{j\mu L}{\sqrt{\left( 1-\chi _{\varphi }\right) }}\right) %
\right] ,  \label{rcq11} \\
&&V_{g}^{\left( 2\right) }\left( 0\right) =2\frac{gm^{2}\mu ^{2}}{8\pi ^{4}%
\sqrt{\left( 1-\chi _{\psi }\right) \left( 1-\chi _{\varphi }\right) }}\left[
\sum_{n=1}^{\infty }f_{\left( 1\right) }\left( \frac{nmL}{\sqrt{\left(
1-\chi _{\psi }\right) }}\right) \right] \left[ \sum_{j=1}^{\infty }\cos
\left( 2\pi j\eta \right) f_{\left( 1\right) }\left( \frac{j\mu L}{\sqrt{%
\left( 1-\chi _{\varphi }\right) }}\right) \right] ,  \label{v2fp} \\
&&V_{2\lambda _{\varphi }}^{\left( 2\right) }\left( 0\right) =\frac{\lambda
_{\varphi }\mu ^{4}}{48\pi ^{4}\left( 1-\chi _{\varphi }\right) }\left[
\sum_{j=1}^{\infty }\cos \left( 2\pi j\eta \right) f_{\left( 1\right)
}\left( \frac{j\mu L}{\sqrt{\left( 1-\chi _{\varphi }\right) }}\right) %
\right] ^{2}.  \label{v2ff}
\end{eqnarray}%
The contribution $V_{\lambda _{\varphi }}^{\left( 2\right) }\left( 0\right) $%
, can be read from the same graph as in Fig.~\ref{fig3}. One constructs the
zeta function $\zeta _{\beta }^{R}\left( s\right) $ from the generalized
zeta function (\ref{rc1.6}), subtracting the divergent part at $s=1$, which
is proportional to $\lambda _{\varphi }$. The factor of $2$ is
to remind that we have to account for two components of the the complex
field, that is, $\varphi _{1}$ and $\varphi _{2}$ which give rise to equal
contributions.\newline
\begin{figure}
\centering
\includegraphics[scale=0.8]{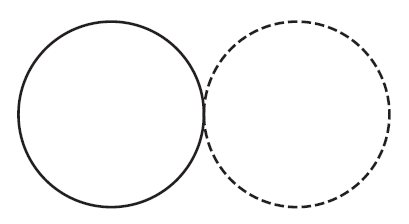}
\caption{Feynman graph representing the contribution from the
self-interaction potential at $\Psi =0$.}
\label{fig4}
\end{figure}
The next contribution $V_{g}^{\left( 2\right) }\left( 0\right) $, Eq.~(\ref%
{v2fp}), is from the interaction between the fields. This correction is
inferred from the graph presented in Fig.~\ref{fig4}, and it is proportional
to the coupling constant $g$. Finally, the last contribution comes from the
interaction of the components of the complex field, $V_{2\lambda _{\varphi
}}^{\left( 2\right) }\left( 0\right) $, which is also a self-interaction.
This contribution is also obtained from tha graph in Fig.~\ref{fig4},
considering the solid line as representing the propagator associated with
the field $\varphi _{1}$ and the dashed one associated with the field $%
\varphi _{2}$. Similar to the result presented in Eq.~(\ref{rcq11}), $%
V_{2\lambda _{\varphi }}^{\left( 2\right) }\left( 0\right) $ in Eq.~(\ref%
{v2ff})\ is proportional to the coupling constant $\lambda _{\varphi }$.

Collecting all results obtained in Eqs~(\ref{rcq9}), (\ref{v2fp}), (\ref%
{rcq11}) and (\ref{v2ff}), one can write the two-loop correction to the
effective potential, at the vacuum state $\Psi =0$, as%
\begin{eqnarray}
\Delta \mathcal{E}_{0z} &=&\left. V^{\left( 2\right) }\left( \Psi
\right) \right\vert _{\Psi =0}  \notag \\
&=&\frac{\lambda _{\psi }m^{4}}{32\pi ^{4}\left( 1-\chi _{\psi }\right) }%
\left[ \sum_{n=1}^{\infty }f_{1}\left( \frac{nmL}{\sqrt{1-\chi _{\psi }}}%
\right) \right] ^{2}+\frac{\lambda _{\varphi }\mu ^{4}}{12\pi ^{4}\left(
1-\chi _{\varphi }\right) }\left[ \sum_{j=1}^{\infty }\cos \left( 2\pi j\eta
\right) f_{1}\left( \frac{j\mu L}{\sqrt{1-\chi _{\varphi }}}\right) \right]
^{2}  \notag \\
&&+\frac{gm^{2}\mu ^{2}}{4\pi ^{4}\sqrt{\left( 1-\chi _{\psi }\right) \left(
1-\chi _{\varphi }\right) }}\left[ \sum_{n=1}^{\infty }f_{1}\left( \frac{nmL%
}{\sqrt{1-\chi _{\psi }}}\right) \right] \left[ \sum_{j=1}^{\infty }\cos
\left( 2\pi j\eta \right) f_{1}\left( \frac{j\mu L}{\sqrt{1-\chi _{\varphi }}%
}\right) \right] .  \label{c2l4}
\end{eqnarray}%
The expression above is the correction to the vacuum energy density obtained
in Eq.~(\ref{rc1.15}). Hence, up to second order, the vacuum energy density
gains correction terms proportional to each coupling constant. Each term on
the r.h.s. of the above expression is affected by the Lorentz violation
parameters $\chi _{\psi }$ or $\chi _{\varphi }$, in the form of a
multiplicative factor as well as in the argument of the Macdonald functions.
In the last term, which is proportional to the coupling constant $g$, the
two Lorentz violation parameters are present since this term comes from the
interaction between the two fields.

Moreover, one can consider the vacuum energy density for the massless fields
case. From Eq.~(\ref{c2l4}) taking $m,\mu \rightarrow 0$, it is easy to see
that the correction to the vacuum energy density reads,

\begin{equation}
\Delta \mathcal{E}_{0z}=\frac{\lambda _{\psi }\left( 1-\chi _{\psi
}\right) }{1152L^{4}}+\frac{\lambda _{\varphi }\left( 1-\chi _{\varphi
}\right) }{12L^{4}}\left( \eta ^{2}-\eta +\frac{1}{6}\right) ^{2}+\frac{g%
\sqrt{\left( 1-\chi _{\psi }\right) \left( 1-\chi _{\varphi }\right) }}{%
24L^{4}}\left( \eta ^{2}-\eta +\frac{1}{6}\right) .  \label{c4ims}
\end{equation}%
In the expression above we have used the value $\zeta_{\text{R}} \left(
2\right) =\frac{\pi ^{2}}{6}$ \cite{elizalde1995zeta,elizalde1995ten} as
well as the Bernoulli polynomials presentend in Eq.~(\ref{rc1.15.b}), with $%
B_{2}\left( \theta \right) =\theta ^{2}-\theta +\frac{1}{6}$. This result is the vacuum energy density in the vacuum state $\Psi =0$.
However, the vacuum state considered here is not the only possible state. We
will see that other stable vacuum states are allowed given certain
conditions \cite{PhysRevD.107.125019}.

Before we analyze the possible stable vacuum states and their stabilities,
we shall investigate the generation of topological mass due the
non-triviality of the spacetime, which is affected by the Lorentz symmetry
violation.

\subsection{Topological mass}

Here we investigate the influence of the conditions studied here and the
Lorentz symmetry violation in the topological mass of the real field $\psi $%
, i.e., the generation of the topological mass at one-loop level. Using the
renormalization condition presented in Eq.~(\ref{rc14.3}) with the
renormalized effective potential given in Eq.~(\ref{rcq4}), we obtain the
following expression for the topological mass of the real field,%
\begin{equation}
m_{\text{T}}^{2}=m^{2}\left[ 1+\frac{\lambda _{\psi }}{4\pi ^{2}\sqrt{1-\chi
_{\psi }}}\sum_{n=1}^{\infty }f_{1}\left( \frac{nLm}{\sqrt{1-\chi _{\psi }}}%
\right) +\frac{g}{\pi ^{2}\sqrt{1-\chi _{\varphi }}}\frac{\mu ^{2}}{m^{2}}%
\sum_{j=1}^{\infty }\cos \left( 2\pi j\eta \right) f_{1}\left( \frac{jL\mu }{%
\sqrt{1-\chi _{\varphi }}}\right) \right] .  \label{rcq5}
\end{equation}%
The Lorentz violation parameters are present in the multiplicative factors $%
\left( 1-\chi _{\psi }\right) ^{-\frac{1}{2}}$ and $\left( 1-\chi _{\varphi
}\right) ^{-\frac{1}{2}}$, and also in the argument of the Macdonald
functions.\newline
\begin{figure}[tbp]
\includegraphics[width=.5\textwidth]{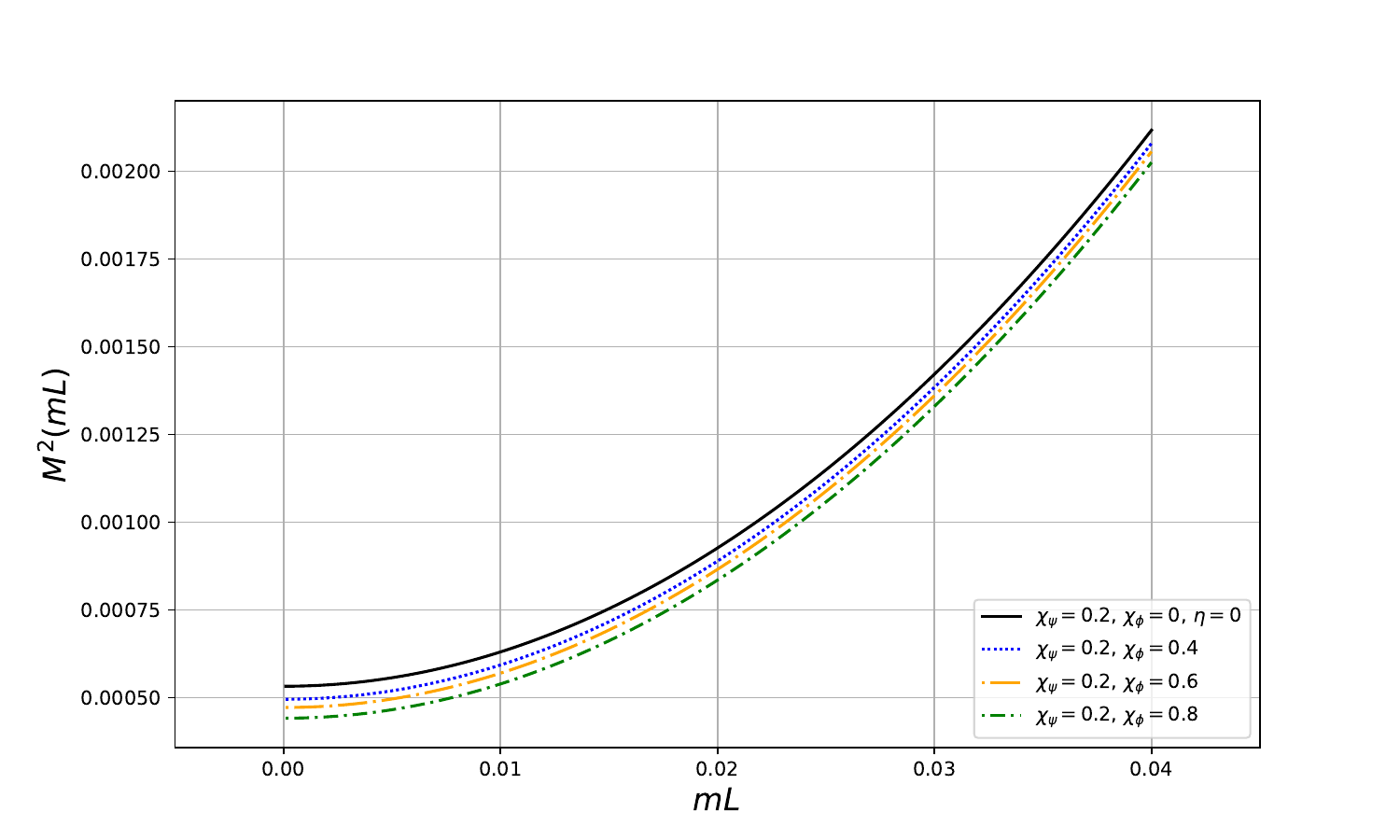}\hfill %
\includegraphics[width=.5\textwidth]{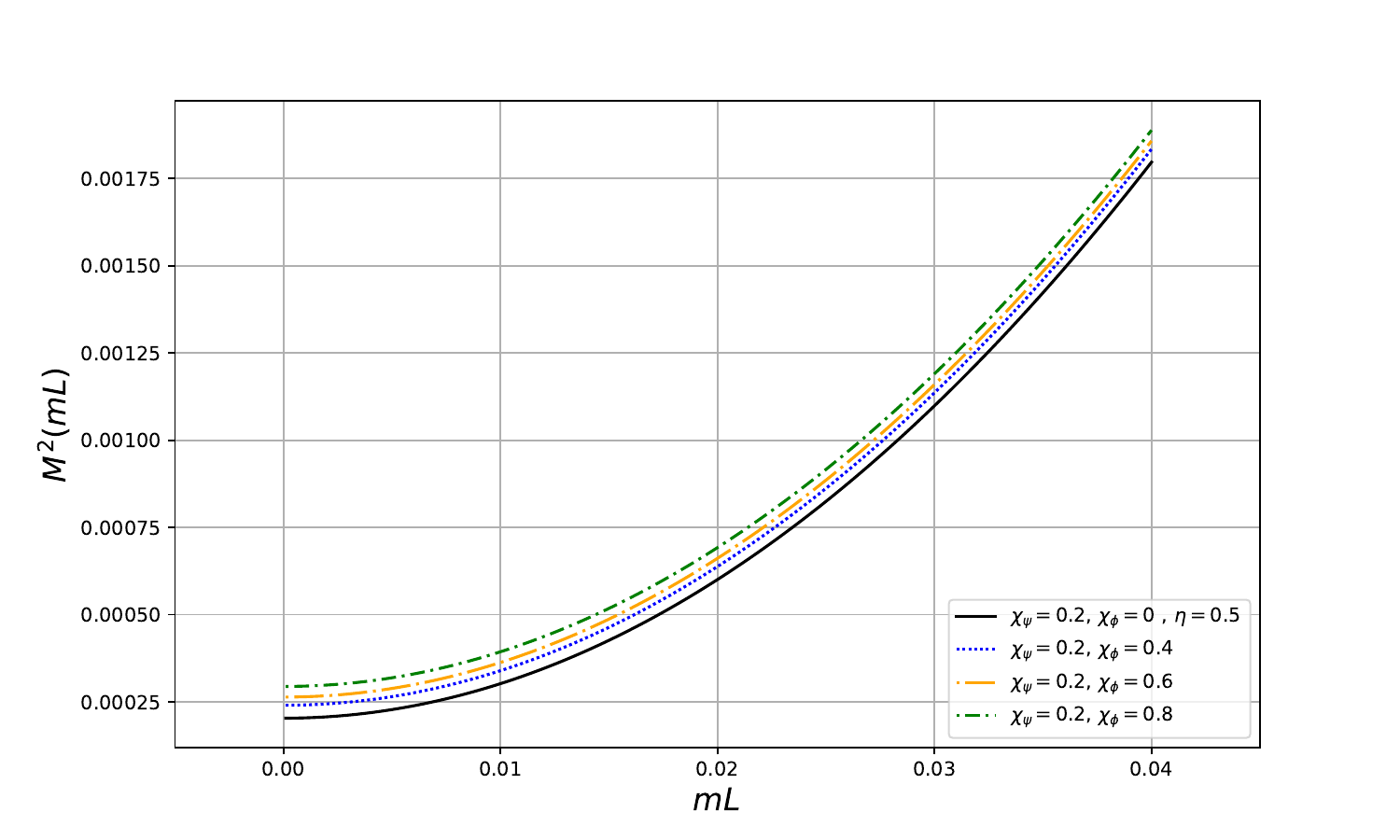}
\caption{Graph of the dimensionless mass squared $M^{2}\left( mL\right)
=L^{2}m_{\text{T}}^{2}\left( mL\right) $, considering $m=\protect\mu $, $%
\protect\lambda _{\protect\psi }=10^{-2}$, $g=10^{-3}$, fixed $\protect\chi %
_{\protect\psi }=0.2$ and different values of $\protect\chi _{\protect\varphi }$
. }
\label{fig5}
\end{figure}
In Fig.~\ref{fig5}, it is shown the dimensionless square mass $M^{2}\left(
mL\right) =L^{2}m_{\text{T}}^{2}\left( mL\right) $ as a function of $mL$,
considering equal masses for the real and complex fields, $m=\mu $, and the
values for the coupling constants $\lambda _{\psi }=10^{-2}$, $g=10^{-3}$,
with fixed $\chi _{\psi }=0.2$ and different values of $\chi _{\varphi }$. The graph
on the left takes the value $\eta =0$ for the quasi-periodic parameter and
shows that as we increase $\chi _{\varphi }$ the topological mass decreases,
while the graph on the right takes the value $\eta =0.5$ and shows the
opposite behavior.

Since the expression (\ref{rcq5}) for the topological mass $m_{\text{T}}^{2}$
does not present any divergence, one can consider the massless field case,
i.e., the limit $m,\mu \rightarrow 0$, and using the same approximation as
before, that is, $K_{\mu }\left( x\right) \approx \frac{\Gamma \left( \mu
\right) }{2}\left( \frac{2}{x}\right) ^{\mu }$ \cite{abramowitz1965handbook}%
, we find the topological mass for the massless field case as,%
\begin{equation}
m_{\text{T}}^{2}=\sqrt{1-\chi _{\psi }}\frac{\lambda _{\psi }}{24L^{2}}+%
\sqrt{1-\chi _{\varphi }}\frac{g}{L^{2}}\left( \eta ^{2}-\eta +\frac{1}{6}%
\right) ,  \label{rcq6}
\end{equation}%
where we have used the result for the Riemann zeta function $\zeta_{\text{R}%
} \left( 2\right) =\frac{\pi ^{2}}{6}$ \cite%
{elizalde1995zeta,elizalde1995ten} and also the Bernoulli polynomials
presentend in Eq.~(\ref{rc1.15.b}), together with the representation $%
B_{2}\left( \theta \right) =\theta ^{2}-\theta +\frac{1}{6}$. As one can
see, the mass correction at one-loop level comes from the self-interaction
term, which is proportional to $\lambda _{\psi }$, and also from the
interaction between the fields, which is proportional to the coupling
constant $g$. Besides, the topological mass has been affected by the Lorentz
symmetry violation, which appears as the multiplicative factors $\sqrt{%
1-\chi _{\psi }}$ and $\sqrt{1-\chi _{\varphi }}$ in the contributions from
the real and complex fields, respectively.

The topological mass presented in Eq.~(\ref{rcq6}) can be positive or
negative, depending on the value of the coupling constants $\lambda _{\psi }$
and $g$, and also on the value of the phase angle, $\eta $, associated with the
quasi-periodic condition imposed on the components of the complex field. If
the conditions are so that the topological mass is negative, this in
principle implicates vacuum instability.\newline
\begin{figure}
\includegraphics[width=.5\textwidth]{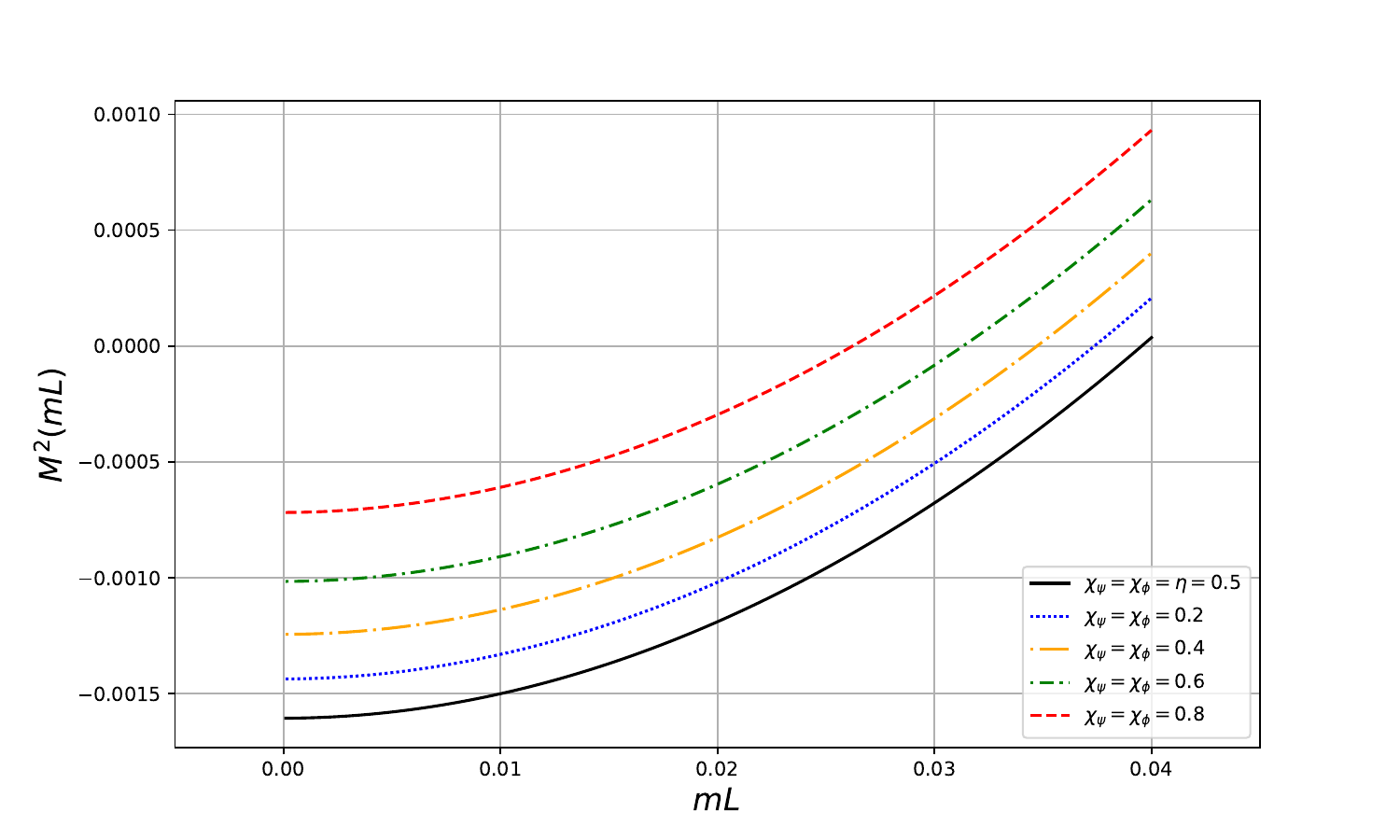}
\caption{Graph of the dimensionless mass squared $M^{2}\left( mL\right)
=L^{2}m_{\text{T}}^{2}\left( mL\right) $, considering $m=\protect\mu $, $%
\protect\lambda _{\protect\psi }=10^{-3}$, $g=10^{-2}$, and equal values for
the Lorentz violation parameters $\protect\chi _{\protect\psi }=\protect\chi %
_{\protect\varphi }$ .}
\label{fig6}
\end{figure}
The graph exhibited in Fig.~\ref{fig6} shows that the
dimensionless squared mass $M^{2}\left( mL\right) $ can be negative, for
some values of the parameters. The graph takes the values $\lambda _{\psi
}=10^{-3}$, $g=10^{-2}$ for the coupling constants, $\eta =0.5$ for the
quasi-periodic parameter and equal values for the Lorentz violation
parameters, that is, $\chi _{\psi }=\chi _{\varphi }$.

In the next section we shall analyze the possible vacuum states and their
corresponding stability conditions, which depends on the values of the
coupling constants and also on the Lorentz violation parameters.

\subsection{Vacuum stability}

\label{sec3.1}In this section we investigate the possible vacuum states and
the stability for the theory described by the action presented in Eq.~(\ref%
{rc2}). However, for simplicity, we consider the massless field case, i.e., $m,\mu
\rightarrow 0$. Hence, for a real field interacting with a complex field,
under periodic and quasi-periodic conditions, respectively, and with the
Lorentz symmetry violation in the $z$-direction, the nonrenormalized
effective potential is,%
\begin{eqnarray}
&&V_{\mathrm{eff}}\left( \Psi \right) _{z}=\frac{\lambda _{\psi }+C}{4!}\Psi
^{4}+\frac{\lambda _{\psi }^{2}\Psi ^{4}}{2^{8}\pi ^{2}\sqrt{1-\chi _{\psi }}%
}\left[ \ln \left( \frac{\frac{1}{2}\lambda _{\psi }\Psi ^{2}}{\nu ^{2}}%
\right) -\frac{3}{2}\right] +\frac{g^{2}\Psi ^{4}}{2^{5}\pi ^{2}\sqrt{1-\chi
_{\varphi }}}\left[ \ln \left( \frac{g\Psi ^{2}}{\nu ^{2}}\right) -\frac{3}{2%
}\right]  \notag \\
&&-\frac{\lambda _{\psi }^{2}\Psi ^{4}}{2^{3}\pi ^{2}\sqrt{1-\chi _{\psi }}}%
\sum_{n=1}^{\infty }f_{2}\left( \frac{n\sqrt{\frac{\lambda _{\psi }}{2}}\Psi
L}{\sqrt{1-\chi _{\psi }}}\right) -\frac{g^{2}\Psi ^{4}}{\pi ^{2}\sqrt{%
1-\chi _{\varphi }}}\sum_{j=1}^{\infty }\cos \left( 2\pi j\eta \right)
f_{2}\left( \frac{j\sqrt{g}\Psi L}{\sqrt{1-\chi _{\varphi }}}\right) .
\label{vs1}
\end{eqnarray}%
Note that in the case under consideration we only need one constant of
renormalization, namely $C$. The renormalization condition which gives the
explicit form of $C$ is presented in Eq.~(\ref{rc14.2}), yielding%
\begin{equation}
C=\frac{3\lambda _{\psi }^{2}}{32\pi ^{2}\sqrt{1-\chi _{\psi }}}\left[ \ln\left( 
\frac{2\nu ^{2}}{\lambda _{\psi }M^{2}}\right)-\frac{8}{3}\right] +\frac{3g^{2}}{%
4\pi ^{2}\sqrt{1-\chi _{\varphi }}}\left[ \ln\left( \frac{\nu ^{2}}{gM^{2}}\right)-\frac{8%
}{3}\right] ,  \label{vs3}
\end{equation}%
where $M$ is a parameter which has dimension of mass. Substituting the
constant $C$ above, on the effective potential (\ref{vs1}), one finds the
renormalized effective potential for the massless field case, i.e.,%
\begin{eqnarray}
&&V_{\text{\textrm{eff}}}^{R}\left( \Psi \right) _{z}=\frac{\lambda _{\psi }%
}{4!}\Psi ^{4}+\frac{\Psi ^{4}}{32\pi ^{2}}\ln \left( \frac{\Psi ^{2}}{M^{2}}%
\right) \left( \frac{\lambda _{\psi }^{2}}{8\sqrt{1-\chi _{\psi }}}+\frac{%
g^{2}}{\sqrt{1-\chi _{\varphi }}}\right) -\frac{25g^{2}\Psi ^{4}}{192\pi ^{2}%
}\left( \frac{\lambda _{\psi }^{2}}{8\sqrt{1-\chi _{\psi }}}+\frac{g^{2}}{%
\sqrt{1-\chi _{\varphi }}}\right)  \notag \\
&&-\frac{\lambda _{\psi }^{2}\Psi ^{4}}{8\pi ^{2}\sqrt{1-\chi _{\psi }}}%
\sum_{n=1}^{\infty }f_{2}\left( \frac{n\sqrt{\frac{1}{2}\lambda _{\psi }}%
\Psi L}{\sqrt{1-\chi _{\psi }}}\right) -\frac{g^{2}\Psi ^{4}}{\pi ^{2}\sqrt{%
1-\chi _{\varphi }}}\sum_{j=1}^{\infty }\cos \left( 2\pi j\eta \right)
f_{2}\left( \frac{j\sqrt{g}\Psi L}{\sqrt{1-\chi _{\varphi }}}\right) .
\label{vs4}
\end{eqnarray}

In order to investigate the vacuum state, up to first order in the coupling
constants $\lambda _{\psi }$ and $g$, we expand the renormalized effective
potential given in Eq.~(\ref{vs4}) in powers of $\lambda _{\psi }$ and $g$ 
\cite{toms1980interacting, PhysRevD.107.125019}, up to first order. The
result is given by%
\begin{eqnarray}
V_{\mathrm{eff}}^{R}\left( \Psi \right) _{z;\lambda _{\psi },g} &=&-\frac{%
\left( 1-\chi _{\psi }\right) ^{\frac{3}{2}}\pi ^{2}}{90L^{4}}+\frac{2\left(
1-\chi _{\varphi }\right) ^{\frac{3}{2}}\pi ^{2}}{3L^{4}}B_{4}\left( \eta
\right) +\frac{\lambda _{\psi }}{4!}\Psi ^{4}  \notag \\
&&+\frac{\Psi ^{2}}{48L^{2}}\left[ \sqrt{1-\chi _{\psi }}\lambda _{\psi }+%
\sqrt{1-\chi _{\varphi }}24gB_{2}\left( \eta \right) \right] ,  \label{vs5}
\end{eqnarray}%
where $B_{4}\left( \eta \right) $ and $B_{2}\left( \eta \right) $ are the
fourth and second order Bernoulli polynominals defined in Eq.~(\ref{rc1.15.b}%
), respectively. The minima of the potential that correspond to the possible
vacuum states are given by%
\begin{equation}
\Psi =0,\qquad\qquad \Psi _{\pm }=\pm \frac{1}{2L}\sqrt{-\left[ \sqrt{%
1-\chi _{\psi }}+\sqrt{1-\chi _{\varphi }}24B_{2}\left( \eta \right) \frac{g%
}{\lambda _{\psi }}\right] }.  \label{vs7}
\end{equation}%
The state $\Psi _{0}$ was considered in the previous section and $\Psi _{\pm
}$ are the other possible vacuum states. Note that the Lorentz violation
parameters direct influence the vacuum states $\Psi _{\pm }$ with the
factors $\sqrt{1-\chi _{\psi }}$ and $\sqrt{1-\chi _{\varphi }}$. Now we
have to analyze which vacuum state is a physical one, that is, which one is
stable. The stability is investigated from the second derivative of the
expanded potential in Eq.~(\ref{vs5}), that is,%
\begin{equation}
\frac{d^{2}}{d\Psi ^{2}}V_{\mathrm{eff}}^{R}\left( \Psi \right) _{z;\lambda
_{\psi },g}=\frac{\lambda _{\psi }}{2}\Psi ^{2}+\frac{1}{24L^{2}}\left[ 
\sqrt{1-\chi _{\psi }}\lambda _{\psi }+\sqrt{1-\chi _{\varphi }}%
24gB_{2}\left( \eta \right) \right] .  \label{vs8}
\end{equation}%
For the vacuum state to be stable, the second derivative of the potential,
evaluated at this state must be greater than zero. This of course depends on
the coupling constants and also on the Lorentz symmetry violation parameters.

We start with the zero vacuum state, that is, $\Psi =0$ which was the
case considered in the previous sections. In this case, from Eq.~(\ref{vs8}%
), one sees that the vacuum state $\Psi=0$ is stable if the following
condition is satisfied,%
\begin{equation}
\lambda _{\psi }>-24R_{\varphi \psi }^{z}gB_{2}\left( \eta \right) ,
\label{vs9}
\end{equation}%
where we have defined the quantity,%
\begin{equation}
R_{\varphi \psi }^{z}=\sqrt{\frac{1-\chi _{\varphi }}{1-\chi _{\psi }}}.
\label{rl}
\end{equation}%
Therefore the Lorentz violation affects the vacuum stability via factor $%
R_{\varphi \psi }^{z}$. However, if the parameters $\chi _{\psi }$ and $\chi
_{\varphi }$ are equal $\chi _{\psi }=\chi _{\psi }$, then $R_{\varphi \psi
}^{z}=1$ and we recover the result presented in \cite{PhysRevD.107.125019}.
The parameter $\eta $ which is associated with the condition on the complex
field, and the coupling constants, $\lambda _{\psi }$ and $g$, also
influence in the vacuum stability. Furthermore, if the coupling constants, $%
\lambda _{\psi }$ and $g$, are taken to be positive along with $R_{\varphi
\psi }^{z}$ and $B_{2}\left( \eta \right) $, then the condition (\ref{vs9})
is automatically satisfied. However, if $B_{2}\left( \eta \right) $ is
negative, the condition (\ref{vs9}) may be violated. Using the explicit form
of the Bernoulli polynomial, that is, $B_{2}\left( \eta \right) =\eta
^{2}-\eta +\frac{1}{6}$, one finds that it is negative for the values of $%
\eta $ in the range,%
\begin{equation}
\frac{1}{2}-\frac{\sqrt{3}}{6}<\eta <\frac{1}{2}+\frac{\sqrt{3}}{6}.
\label{vs10}
\end{equation}%
For $\eta $ inside the above interval, we have to apply condition (\ref{vs9}%
) in order to obtain the allowed values for the coupling constants such that
the vacuum state $\Psi =0$ is stable. For instance, let us consider the
value $\eta =1/2$, i.e., the components of the complex field, $\varphi _{i}$%
, are twisted. Then, the condition of stability written in Eq.~(\ref{vs9}),
becomes,%
\begin{equation}
\lambda _{\psi }>2R_{\varphi \psi }^{z}g.  \label{vs11}
\end{equation}%
The condition above is in agreement with the one found in \cite%
{PhysRevD.107.125019} and also \cite{toms1980interacting} if there was no
Lorentz symmetry violation or the parameters $\chi _{\psi }$ and $\chi
_{\varphi }$ are equal. Since the model includes the Lorentz symmetry
violation and allows different values for the parameters, we have to
consider the quantity $R_{\varphi \psi }^{z}$. The topological mass, for
this case, takes the following form,%
\begin{equation}
m_{\text{T}}^{2}=\frac{\sqrt{1-\chi _{\psi }}\lambda _{\psi }-2\sqrt{1-\chi
_{\varphi }}g}{24L^{2}},  \label{vs12}
\end{equation}%
which agrees with the result obtained in Eq.~(\ref{rcq6}) with $\eta =1/2$.
Note also that if the interaction between the fields is not present, that
is, $g=0$, and also there is no Lorentz symmetry violation, then it
reproduces previous known results \cite{Ford:1978ku}.

Next we consider the other vacuum states, i.e., $\Psi =\Psi _{\pm }$ from
Eq.~(\ref{vs7}). In this case, the second derivative of the effective
potential, Eq.~(\ref{vs8}), give rise to the following condition:%
\begin{equation}
\lambda _{\psi }<-24R_{\varphi \psi }^{z}gB_{2}\left( \eta \right) .
\label{vs13}
\end{equation}%
Note that the states $\Psi _{\pm }$ depends on the Lorentz symmetry
violation parameters $\chi _{\psi }$. However, the stability condition is
affected if and only if the parameters are different, $\chi _{\psi }\neq
\chi _{\varphi }$. Since we are taking the coupling constants as positive
numbers, then $B_{2}\left( \eta \right) <0$ which gives the same interval as
the Eq.~(\ref{vs10}) for the parameter $\eta $. From this observation, one
can conclude that the possible values of the parameter $\eta $, that makes
possible to $\Psi _{\pm }$ be a stable vacuum state are within the interval
given in Eq.~(\ref{vs10}). Considering again the value $\eta =1/2$, we
obtain the stability condition as,%
\begin{equation}
\lambda _{\psi }<2R_{\varphi \psi }^{z}g.  \label{vs14}
\end{equation}%
Furthermore the topological mass now reads,%
\begin{equation}
m_{\text{T}}^{2}=\frac{2g\sqrt{1-\chi _{\varphi }}-\lambda _{\psi }\sqrt{%
1-\chi _{\psi }}}{12L^{2}}.  \label{vs15}
\end{equation}%
Since we are taken a different vacuum state from $\Psi=0$, we expect that
the topological mass (\ref{vs15}) differ from the one given in Eq.~(\ref%
{vs12}), which is the case here.

The vacuum energy density is also considered in the study of the vacuum
stability. Considering the case of the vacuum state being $\Psi =0$
and in addition, the value $\eta =1/2$ for the parameter of the
quasi-periodic condition, one obtains the vacuum energy density as being the
same as the one presented in Eq.~(\ref{rc1.16}) with $\eta =1/2$. However,
considering that the vacuum state is $\Psi =\Psi _{\pm }$, (\ref{vs7}), one
can substitute this state in the expanded effective potential given in Eq~(%
\ref{vs5}), to obtain the following expressions for the vacuum energy
density,%
\begin{equation}
V_{\mathrm{eff}}^{R}\left( \Psi _{\pm }\right) _{z;\lambda _{\psi },g}\simeq
-\frac{\left( 1-\chi _{\psi }\right) ^{\frac{3}{2}}\pi ^{2}}{90L^{4}}+\frac{%
2\left( 1-\chi _{\varphi }\right) ^{\frac{3}{2}}\pi ^{2}}{3L^{4}}B_{4}\left(
\eta \right) -\frac{1}{384\lambda _{\psi }L^{4}}\left[ \sqrt{1-\chi _{\psi }}%
\lambda _{\psi }+\sqrt{1-\chi _{\varphi }}24gB_{2}\left( \eta \right) \right]
^{2}.  \label{vs16}
\end{equation}%
Note that the first two terms on the r.h.s. of Eq.~(\ref{vs16}) are in
agreement with the vacuum energy density presented in Eq.~(\ref{rc1.16}).
However, the last term presents a dependence on the coupling constants $%
\lambda _{\psi }$ and $g$, which does not appears in the case where the
vacuum state is $\Psi =0$. It is important to point out that the
expression above is only an approximation, since we are taking the expansion
of the effective potential presented in Eq.~(\ref{vs5}).

From Eqs. (\ref{vs9}) and (\ref{vs13}), we conclude that the stability of
the vacuum state is determined by the values of the Lorentz violation
parameters $\chi _{\psi }$ and $\chi _{\varphi }$, coupling constants $%
\lambda _{\psi }$ and $g$, and also depends on the value of the parameter $%
\eta $ of the condition for the complex field. However it shows no
dependence on the parameter $L$ \cite{PhysRevD.107.125019}.

To end this section let us discuss the effect of the Lorentz violation
scenario in which we are considering in the $\tau $, $x$ and $y$ directions.
In the temporal direction, the effect of the Lorentz violation on the vacuum
energy density and topological mass of each field is only by means of a
global multiplicative factor $\sqrt{1+\chi}$ in the denominator. In
contrast, in the spatial directions $x$ and $y$, the effect of the Lorentz
violation is only by means of a global multiplicative factor $\sqrt{1-\chi}$
in the numerator. In these cases, the vacuum energy density and topological
mass are trivially attenuated by the mentioned factors, since $\chi$ is to be
considered smaller than unit (see Refs. \cite{cruz2020casimir, Aj}).
However, the Lorentz violation effect is nontrivial when we consider the same
direction as the one along which the condition is imposed, as our study has indicated.

\section{Concluding remarks}

\label{sec5}

The vacuum energy density, its loop correction and the generation of the
topological mass was investigated for a system consisting of a real field,
interacting with a complex one via quartic interaction. Besides, the
self-interaction potential of each field was included. The whole system is
considered in a scenario in which the CPT-even aether-type violation of
Lorentz symmetry takes place. Both vacuum energy density and topological mass arise
from the nontrivial topology of the spacetime, induced by periodic and
quasi-periodic conditions imposed on the fields. The real field is subject
to periodic condition while the complex is assumed to obey the
quasi-periodic condition.

Considering that the CPT-even aether-type violation of Lorentz symmetry
occurs in the $z$-direction, the vacuum energy density was obtained in Eq.~(%
\ref{rc1.15}), for massive fields, and in Eq.~(\ref{rc1.16}), for the
massless fields case. These results shows the dependence of the vacuum
energy density on the lenght $L$ and also in the parameters of the Lorentz
symmetry violation $\chi _{\psi }$ and $\chi _{\varphi }$ of the real and
complex fields, respectively. The two-loop correction contributions to the
vacuum energy density, considering both massive and massless fields cases,
have been presented, respectively, in Eqs.~(\ref{c2l4}) and (\ref{c4ims}),
which turn out to be proportional to the coupling constants $\lambda _{\psi
} $, $\lambda _{\varphi }$ and $g$. Also, the topological mass for this
system was obtained in Eqs.~(\ref{rcq5}) and Eq.~(\ref{rcq6}), for the
massive and massless cases, respectively. Furthermore, it was investigated
the possible vacuum states and their corresponding stability conditions. The
possible vacuum states have been presented in Eq.~(\ref{vs7}) and the
corresponding stability for each state is written in Eqs.~(\ref{vs9}) and (%
\ref{vs13}), which exhibit a dependence on the values of the coupling
constantes $\lambda _{\psi }$, $\lambda _{\varphi }$, $g$ and on the
parameter $\eta $ of the quasi-periodic condition, it also depends on the
parameters of the Lorentz symmetry violation $\chi _{\psi }$ and $\chi
_{\varphi }$. It is shown that although the possible vacuum states depend on
the parameters $\chi _{\psi }$ and $\chi _{\varphi }$, if $\chi _{\psi }$
and $\chi _{\varphi }$ are equal, the stability condition does not depend on
theses parameters and we recover the case in which there is no Lorentz
symmetry violation \cite{PhysRevD.107.125019}, for the stability condition.
Also, the vacuum stability does not depend on the periodic parameter $L$.

We have also discussed the rather trivial effects of the Lorentz symmetry
violation on the vacuum energy density and topological mass in the $\tau$, $%
x $ and $y$ directions. As we have pointed out, this happens by means of
global multiplicative factors, that is, $\sqrt{1+\chi}$ in the denominator
in the case of temporal direction, and $\sqrt{1-\chi}$ in the numerator in
the case of $x$ and $y$ directions. However, we have seen that a nontrivial effect
takes place in the $z$-direction in the massive case, where a periodic and quasi-periodic conditions are also considered.

Finally, the toy model that has been considered here helps us to analyze some important aspects that may be present in a more realistic model
as, for instance, additional contributions to the free vacuum energy density arising due to the interactions, and the necessary stability requirements associated with different vacua inherent to 
the model. How this stability is reached is a key aspect of the theory for the calculation of the vacuum energy density and the topological mass. The inclusion of interaction terms, may also be considered as a small additional step toward more realistic situations considering fermion and gauge fields. The Lorentz violation term considered comes as an extra ingredient that modifies the results. The modifications, in an improved model, can offer us a way to do phenomenology associating the results with experimental data and, as a consequence, providing a constraint on the Lorentz violation parameter, similar to what has been done in Ref. \cite{deFarias:2023xjf}.

{\acknowledgments} The author H.F.S.M. is partially supported by the
Brazilian agency CNPq under Grant No. 308049/2023-3.

%\bibliographystyle{JHEP}
%\bibliography{ref1}

\end{document}